# Rational design of carbon-based materials for purification and storage of energy carrier gases of methane and hydrogen

Shohreh Mirzaei[1,2], Ali Ahmadpour[1,2,*], Zongping Shao[3], Arash Arami-Niya[3,**]

[1] Department of Chemical Engineering, Faculty of Engineering, Ferdowsi University of Mashhad (FUM), P.O. Box 91779-48944, Mashhad, Iran.

[2] Industrial Catalysts and Adsorbents and Environment Research Lab., Oil and Gas Research Institute, FUM, Iran.

[3] Discipline of Chemical Engineering, Western Australian School of Mines: Minerals, Energy and Chemical Engineering, Curtin University, GPO Box U1987, Perth, WA 6845, Australia.

## Abstract

Today, fast-growing energy demands and fuel resource depletion are among the hottest concerning issues that treating our world. So, a huge need is felt to find efficient, affordable and eco-friendly energy storage and production systems. Much current research effort proved that gaseous energy carriers such as $CH_4$ and $H_2$ seem to be the right choice for alternative fuel resources. However, the most important challenge with this new-faced resource is the comparatively low volumetric energy storage density. Fortunately, the high-pressure gas storage technique inside the porous media of solid adsorbent is considered as one best way to tackle the energy density problem. Famous family of porous carbon materials, with a suitable pore size distribution centred in the micropore range and a large number of adsorption sites per volume of solid, open up a great scope for gas storing applications. This review article represents the state-of-the-art with a precise focus on what has and can be done to improve/enhance the gas/energy storage capacity of traditional and novel structures of low-cost carbon-based adsorbents. We review a wide variety of design strategies to synthesis carbonaceous adsorbents, with a strong focus on creating the connection between structural properties and gas adsorption performance. In this regard, various synthesis techniques have been studied with emphasis on the more interesting recent progress that allows better control and optimisation of porosity of porous carbons for maxing gas storage capacity. We will also show that carbon-based adsorbents, particularly activated carbons, have been extensively studied and remain a powerful candidate in the search for an energy carrier economy. In the end, a perspective is provided to forecast the future development of carbon-based materials.

* Corresponding author. Tel: +98-51-38805006. Email: Ahmadpour@um.ac.ir
** Corresponding author. Tel: +61-8-9266-5482. Email: Arash.araminiya@curtin.edu.au

Contents

## 1. Introduction

The energy crisis and environmental effects are interconnected concerning issues the world is dealing with today. For years, the nations have been alarmed by the possible climate changes due to an increasing level of carbon dioxide in the atmosphere. Moreover, according to the worldwide energy consumption statistics, oil and coal as the primary available energy resources will deplete in the next few decades (Figure 1) [1, 2]. Therefore, a significant challenge is to find efficient potential energy sources to replace conventional fossil fuels. Gasses such as methane ($CH_4$) and hydrogen ($H_2$) are the best alternative green energy carriers. The combustion of these gases generates a significant amount of energy (Methane 50-55 $MJ.Kg^{-1}$ and hydrogen 120-142 $MJ.Kg^{-1}$) and, at the same time, produces a clean residue [3, 4]. It has been estimated that these energy carriers can supply 80% of the energy demands [5].

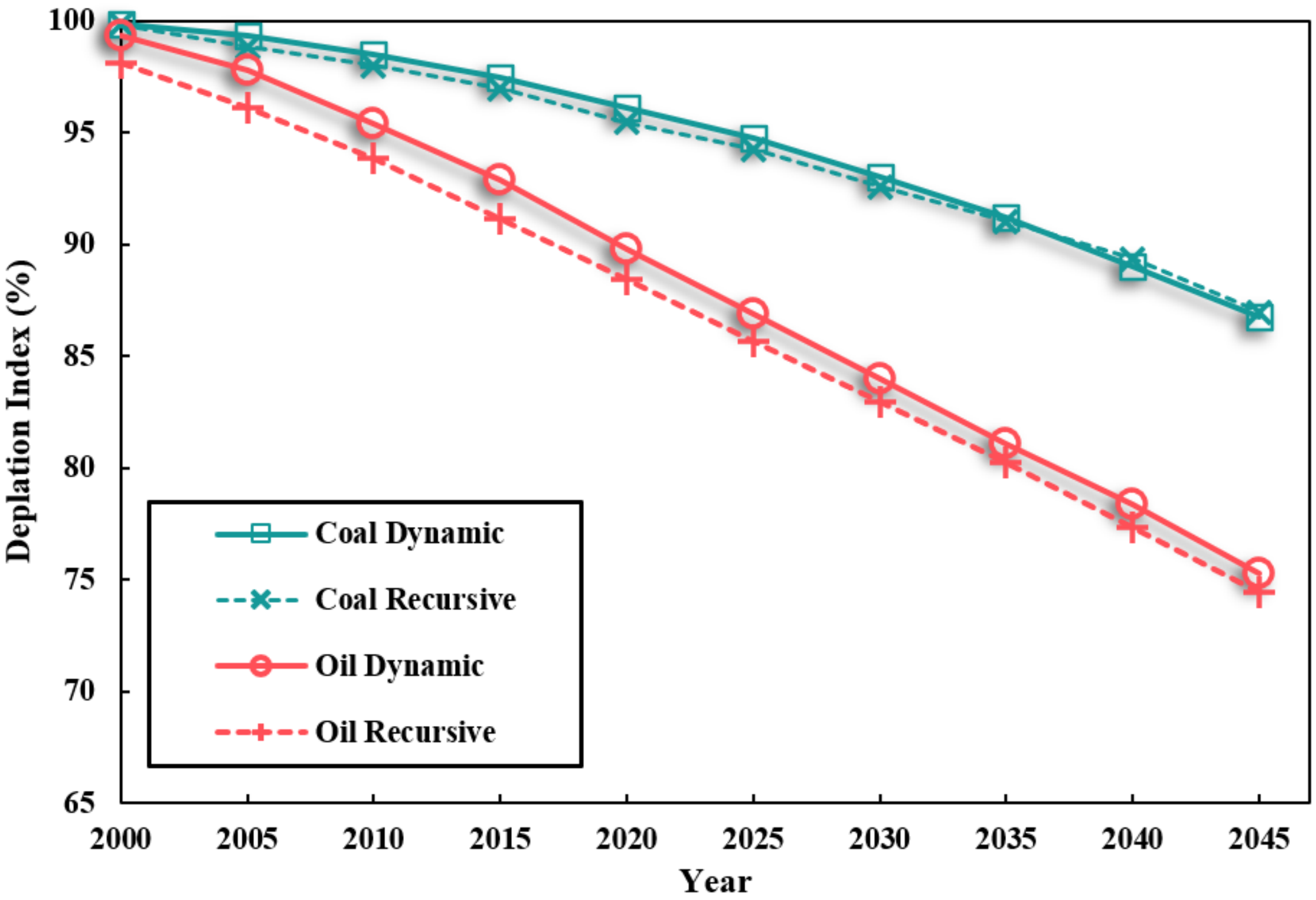

**Figure 1.** Global depletion of coal and oil as the primary resources of fossil fuel predicted by recursive-dynamic MIT Emissions model [6].

Methane, the main constituent of natural gas (NG), is known as an alternative energy source due to its accessibility and clean-burning properties over conventional fuels. Among all hydrocarbons, methane emits the least amount of pollutants during the combustion [5]. The U.S. Energy Information Administration (EIA) revealed that "natural gas expected to remain most-consumed fuel in U.S. industrial sector up to 2050" [7]. Also, EIA projects that natural gas consumption in Asia will continue to outpace supply (see Figure 2) [8]. The agency reported that natural gas production will continue growing through 2050 amid rising global energy consumption, especially in power generation and transportation [7]. A vehicle running on $CH_4$ would produce 70% less CO, 87% less non-methane organic gas, 87% less nitrogen oxide, and 20% less $CO_2$ compared with those running on gasoline [9]. However, the low energy yield of each litre of natural gas (0.033 MJ) at the standard condition in contrast with the gasoline (34.2 MJ) is one of the main drawbacks in the NG application as a source of energy.

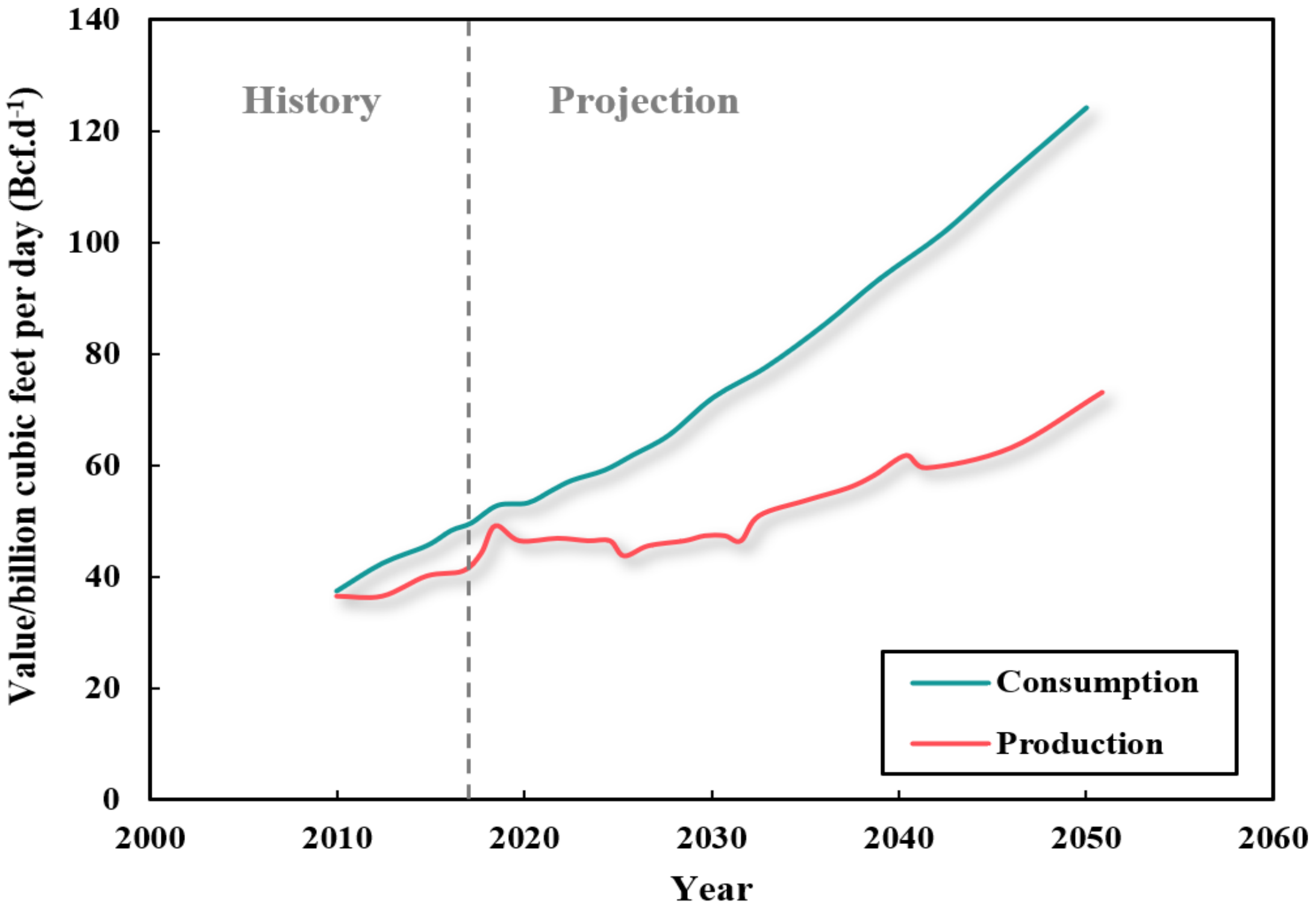

**Figure 2.** Asia natural gas consumption and natural gas production were specified from 2010 and predicted up to 2050 [8].

Natural gas storage has predominately been conducted through compressed natural gas (CNG) or liquefied natural gas (LNG) systems. However, these methods have been considered expensive[1] due to the high compression and liquefaction prices and the costly storage tanks that should withstand high pressures (200-250 bar) or low temperatures (-162 °C). Adsorbed natural gas (ANG) systems have been proposed to increase natural gas's energy density at low pressures (35-65 bar) and room temperature. To be economically equivalent to the other NG storage techniques, the US Department of Energy (DOE) initiated the $CH_4$ storage targets; in 1993, the DOE defined the storage target 150 $cm^3.cm^{-3}$, volume of $CH_4$ at standard condition per volume adsorbent ($V.V^{-1}$), at 25 °C and 35 bar pressure. The DOE target was raised up to 180 $cm^3.cm^{-3}$ after a few years and then recently to 350 $cm^3.cm^{-3}$ (for volumetric measurements) and 0.5 $kg.kg^{-1}$ (for gravimetric capacity)[12, 13].

Besides $CH_4$, the use of $H_2$, as a clean source of energy with zero carbon foot print, attracted considerable interest in recent years due to its fascinating characteristics; $H_2$ as the light element (second lightest element after helium), produces the largest amount of energy on a gravimetric basis in comparison with other fuels as shown in Figure 3 [5]. On a mass basis, the energy content of $H_2$ (120 $MJ.kg^{-1}$) is nearly three times larger than gasoline (44 $MJ.kg^{-1}$). Besides, as a fuel, $H_2$ produces water after the combustion reaction. However, $H_2$ has a low volumetric energy density at the ambient temperature and atmospheric pressure. Liquified $H_2$ has a relatively higher energy density of 8 MJ.l$^{-1}$, but still is lower than the conventional fuels such as gasoline (32 MJ.l$^{-1}$). Developments in $H_2$ storage technologies, such as metal hydrides, spillover, physisorption and chemisorption, might improve the $H_2$ low volumetric energy density. However, most of these strategies are still far from commercialization due to economic and/or technical challenges. Among the current technics, physisorption of $H_2$ on porous materials seems to be promising due to its fast kinetics and the reversibility of the storage process [14]. To make the adsorptive storage of $H_2$ comparable to liquefied $H_2$, in 2020, the US Department of Energy (DOE) specified a target adsorbed $H_2$ value of 0.040 kg $H_2$.l$^{-1}$ and 0.055 kg $H_2$.kg$^{-1}$ system on a volumetric and gravimetric basis (See Figure 3).

[1] **A)** The study on the economic feasibility associated with compressed natural gas vehicle fueling provide estimated cost range of $250000-$600000 for instalation of medium CNG station with capacity of 500-800 gasoline gallon equivalent per day.
[10] M. Smith, J. Gonzales, Costs associated with compressed natural gas vehicle fueling infrastructure, National Renewable Energy Lab.(NREL), Golden, CO (United States), 2014..
**B)** The expenses of diffrent phase related to the LNG opetation are generally included exploration & production ($0.60–1.2 per one million British thermal units, or MMBtu), liquefaction ($0.90–1.30/MMBtu), shipping ($0.50–1.80/MMBtu), storage & regasification ($0.40–0.60/MMBtu).
[11] J. Cho, G.J. Lim, S.J. Kim, T. Biobaku, Liquefied natural gas inventory routing problem under uncertain weather conditions, International Journal of Production Economics 204 (2018) 18-29.

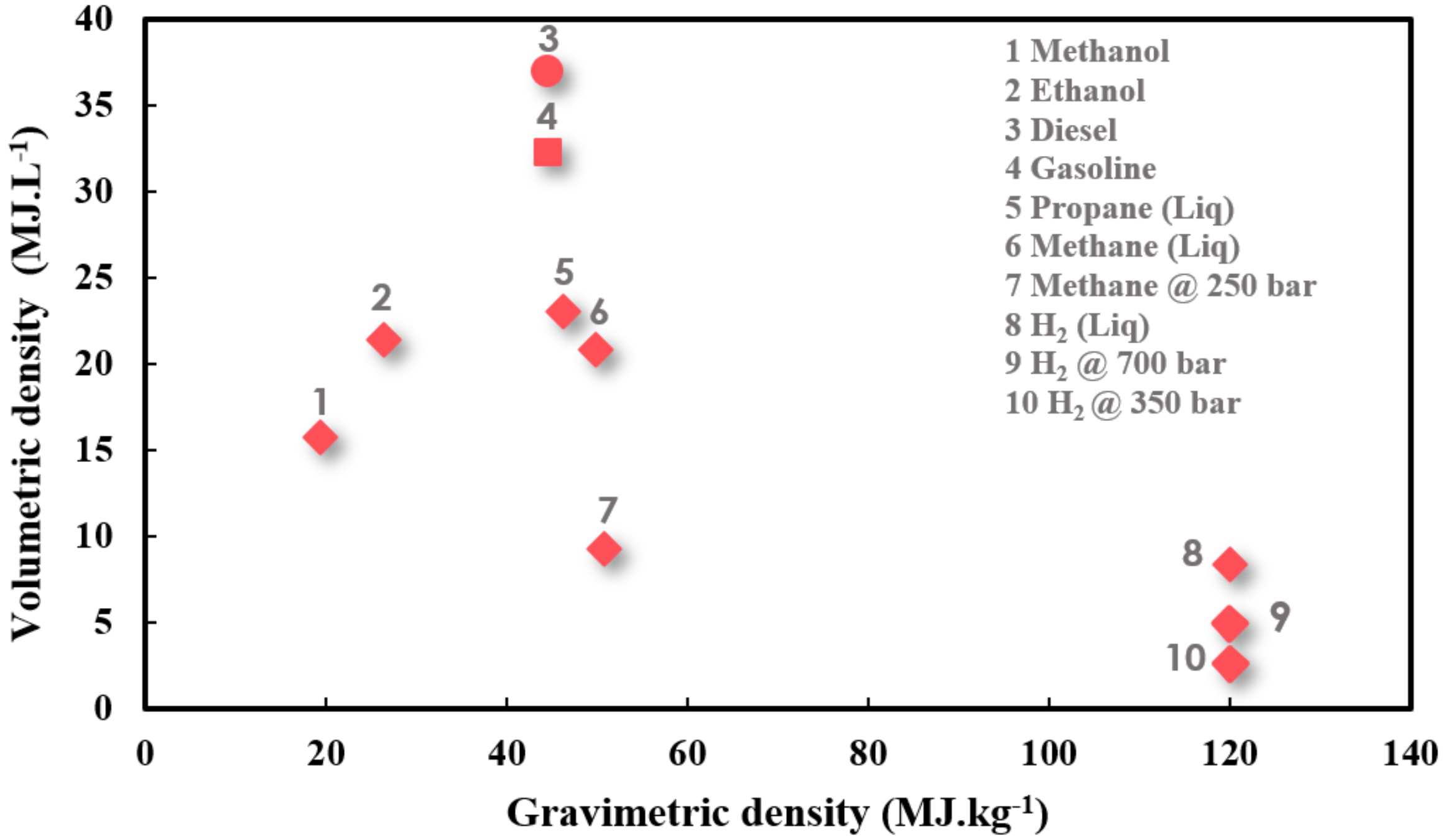

**Figure 3.** Comparison of energy density for different types of fuels on gravimetric and volumetric scales [5].

Several types of adsorbents have been investigated in the literature for gas storage applications, including porous carbons [15], zeolites [16], metallic organic frameworks (MOFs) [17], porous aromatic frameworks (PAFs) [12, 18] and metal oxides [19, 20]. Based on what has been reported, to maximise the gas storage capacity, adsorbents should meet the following requirements:

(1) micropore size distribution with predominant pores around 0.8 -1.1 nm width for methane and 0.6 nm for hydrogen storage (should be noted that with such an optimum PSD, two $CH_4$ and $H_2$ molecules can be adsorbed easily inside the micropores)[21-23],

(2) high packing density to guarantee that volumetric storage capacity will be high [24-27],

(3) appropriate thermal properties (i.e. high specific heat capacity and low heat of adsorption) to minimize the thermal fluctuations inside the adsorbent bed [28, 29],

(4) better to be extremely hydrophobic [23, 30],

(5) preferably inexpensive and potential for mass production [31, 32].

The advantages and disadvantages of some groups of adsorbents are listed in Table 1. As can be seen, although some of the novel adsorbents such as MOFs or PAFs can be designed to posss optimum pore size distribution and an exceptional gas adsorption capacity (near the latest DOE target), their costly and complex synthesis methods hinders their practical application.

Carbon-derived adsorbents are one of the main studied classes of adsorbents for gas storage and energy-related applications due to their particular characteristics such as high porosity, and chemical/thermal/mechanical stability. More importantly, compared to other porous materials that require the use of expensive precursors and time-consuming synthesis, production of carbon derived adsorbents is affordable. As shown in Figure 4, there is a wide range of carbon-derived products, including activated carbons, carbon fibres, carbon foams, carbon molecular sieves and carbon nano-allotropes (i.e. fullerene, carbon nanotube, graphene sheets and carbon graphitic structures).

**Table 1.** Some critical features of well-known groups of adsorbents.

| Adsorbent's type | Feature | |
|---|---|---|
| | **Advantage** | **Disadvantage** |
| **Carbons** [15] | + Large surface area (up to 3000 $m^2.g^{-1}$), + Tunable pore size distribution,<br>+ Light weight<br>+ High adsorptive characteristics ($CH_4$ uptake> 190 $V.V^{-1}$ & $H_2$ uptake ~ 11.5 wt.%)<br>+ Abundant and eco-friendly precursors<br>+ Simple production methods<br>+Possiblity structural modifications<br>+ Excellent stability in extreme conditions of temperature and humidity<br>+ Generally hydrophobic surfaces<br>+ High heat and electrical conductivities<br>+ Acceptable cost, the capacity of large production<br>+ Excellent mechanical properties | - Low adsorption capacity and selectivity in comparison with novel adsorbents such as MOFs.<br>- Challenging to prepare different batches of a carbonaceous adsorbent with a uniform structure (As zeolites or MOFs). |
| **Zeolites** [16] | + Medium gas storage capacity<br>+ Large portipon of micro/mesopores<br>+ Low cost<br>+ Good mechanical properties | - Relatively low surface area in comparison with ACs and some novel adsorbents<br>- High affinity to water<br>- Difficult synthesis process<br>- Regeneration needs more effort in comparison with other classes of the adsorbents |
| **Metallic organic frameworks (MOFs)** [17] | + High surface area (area up to 7000 $m^2.g^{-1}$)<br>+ Tunable pore volume/size distribution<br>+ Higher gas storage capacity usually more than carbon materials[33] ($CH_4$ uptake> 250 $V.V^{-1}$ & $H_2$ uptake ~11 wt.%)<br>+ High gas selectivity | - Expensive raw material and production methods<br>- Difficult synthesizing processes<br>- Low thermal stability<br>- Poor mechanical strength<br>- Unstable in the presence of moisture |
| **Porous aromatic frameworks (PAFs)** [12, 18] | + Very high surface area (up to 7500 $m^2.g^{-1}$)<br>+Very high gas storage/separation capacity ($CH_4$ uptake> 250 $V.V^{-1}$ & $H_2$ uptake ~ 64 wt.% )<br>+ Exceptional thermal and chemical stability<br>+ Synthesised via various coupling reactions<br>+ Chemically resistant to harshly acidic, alkaline and humid environments | - Current coupling reactions for PAF syntheses are expensive<br>- Organometallic catalysts which are used in PAFs preparation method hardly can be removed from the framework<br>- Difficulties in morphologies engineering and shapes for specific applications |
| **Metal Oxides** [19, 20] | + Good electrical and ionic conductivity properties<br>+ Ability of reversible capacity<br>+ High mechanical stability<br>+ Relatively high gravimetric $H_2$ storage capacity | - Expensive materials<br>- Slow reaction Kinetics<br>- High dissociation temperatures<br>- Unpublished data for $CH_4$ uptake over last years |

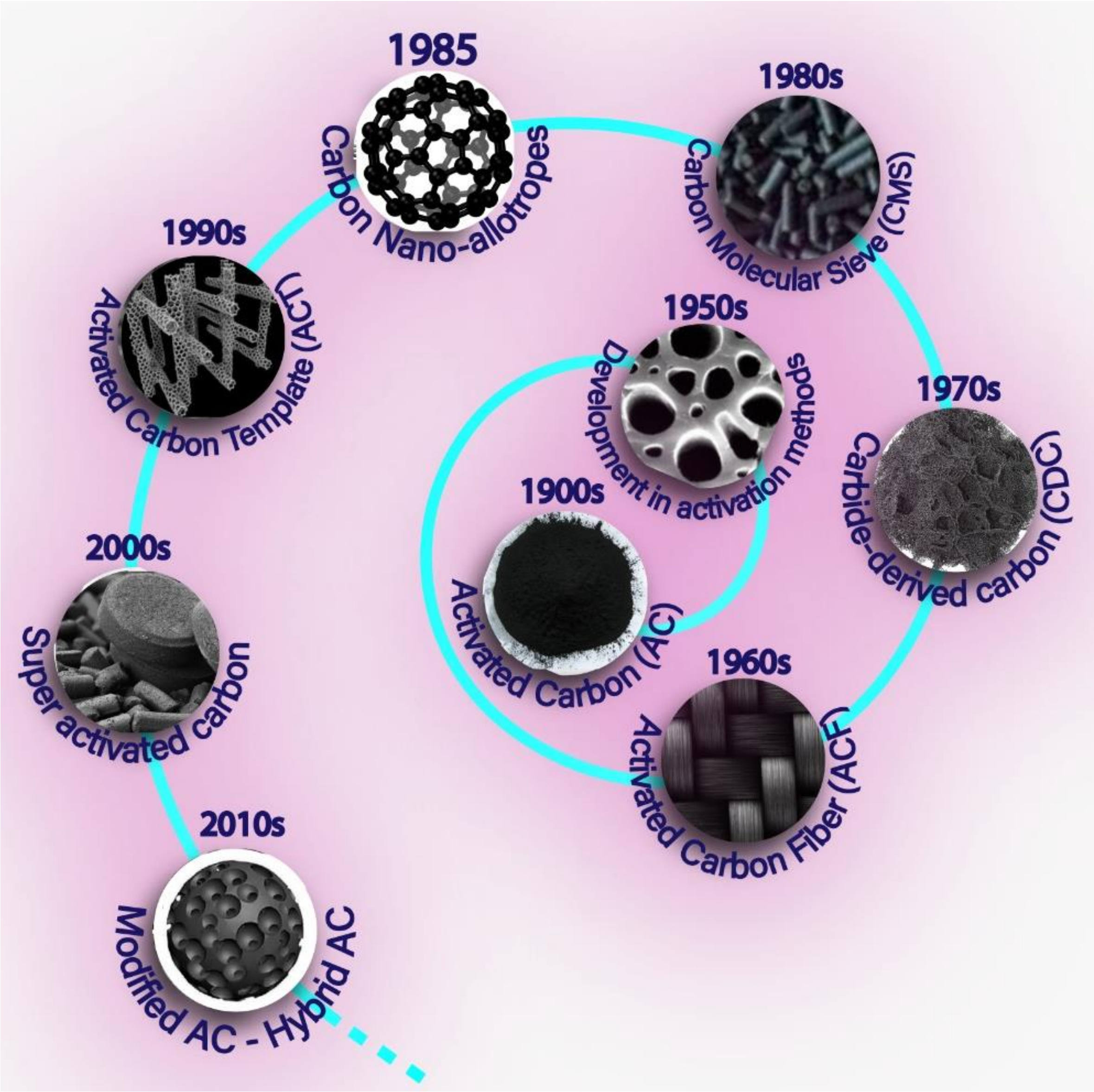

**Figure 4.** A road map sketch over the developmental history of the porous carbons

The gas adsorption capacity of solids strongly depends on their porosity and surface properties. Careful fabrication of carbonaceous matters can significantly improve their properties for specific applications. For instance, some new-reported activated carbons possess a significant porosity and surface area of more than 3000 $m^2.g^{-1}$. However, it needs to be considered that the porosity of adsorbents does not necessarily guarantee their high gas storage capacity. Numerous investigations are still going on to optimise production methods for the cost-effective microporous carbons that satisfy the latest value of the DOE target. The linear relationship between methane adsorption capacity and micro/meso porosity of adsorbents has been discussed by Rodríguez-Reinoso et al. and Monge et. al [31, 34]. They found out that the sample with the highest BET surface area (Brunauer, Emmett and Teller, $S_{BET}$) shows the largest gravimetric $CH_4$ uptake at any desired pressure. However, the low packing density of this sample directly affects its volumetric storage capacity, which results in its relatively low volumetric $CH_4$ uptake. On the other hand, the sample with a medium level of microporosity and packing density shows the highest value of volumetric $CH_4$ capacity among all the prepared ACs. It can be concluded that the best sample for ANG purposes is the one with the best balance among mico porosity and the packing density[13, 34].

The majority of available research outcomes on carbon preparation rely on a trial and error approach with relatively considerable uncertainty and poor reproducibility of the results. One best way to maximise carbon-based materials' gas storage performance is to design them with a precise and predictable synthesis approach. Such an approach provides a specific large-scale synthesis pathway which is a complete departure from conventional methods. This review aims to concentrate on five topics of; **(1)** theoretical and experimental design of different classes of carbon-based adsorbents for gas storage/separation applications with deep emphasis on $H_2$ and $CH_4$, **(2)** conventional carbon-based adsorbents **(3)** novel structured microporous carbons, **(4)** carbon nano allotropes**, (5)** the future for carbon-based gas adsorbents.

In the present review study, we mostly presented more recent research articles though the previous related literature were also monitored, and used if it was necessary. According to statistics, up to January 2022, about 5,214 and 13,701 publications were respectively matched with search topics of "carbon+methane adsorption/storage/separation" and "carbon+hydrogen dsorption/storage/separation" [35]. Within the last 10 years, the number of publications in this particular field has increased and this is reflecting the rapid expansion of this research topic [35]. Therefore, we believe it is an appropriate time to provide a comprehensive guide to the developments in this field over the past decade.

## 2. Activated carbons

Activated carbons (ACs) is one the main members of porous carbon materials which have been extensively investigated in different fields of application. Their tunable micro/mesoporosity has been instrumental in achieving their best performances, especially for fuel gas storage and separation requests. A variety of synthesis approaches have been developed to control the pore size distributions in these materials - some of these were undoubtedly succeeded in theoretical and experimental assessment [23, 24, 30, 32, 36-42]. In this work, some of the efficient practices reported in the literature were reviewed.

ACs are prepared through the decomposition of organic substances with high carbon content (which is called carbonization process) followed by the activation of the carbonized materials to improve the porosity, as presented in Figure 5. The starting materials for AC production could be classified into two main groups of non-renewable materials and renewable biomass resources. As shown in Table 2, various fossil-based precursors such as coal [24, 43], petrol coke [44, 45], pitch [42, 46] as well as synthetic polymers [47, 48] are categorised as non-renewable precursors. Renewable sources with inherent advantages like cost-effectiveness and availability are classified into three main groups including (i) polysaccharides (e.g. starch [49] and chitosan [50]) (ii) raw or waste biomass (e.g. date seed [51] and Mandarin peels [52]) and (iii) microorganisms, (e.g. microalgae [53] and fungi [14]) (see Table (3)).

According to the literature, the precursors with a high carbon content are promising candidates for the gas storage/separation application [26, 32, 39]. However, it is possible to enhance the fixed carbon content of the starting materials by prewashing them in two different media of water and/or mild acid [54]. The unwanted organic attachments/contaminations like sands and loose ash can be eliminated by water washing. The chemically bonded compound including ash elements can be effectively removed from the solid structure by the acid washing process. Suppose the washed-precursors still possess a low amount of carbon, it is essential to raise it by removing light volatiles through an initial heat treatment (i.e. pre-carbonisation) before the production procedure[54].

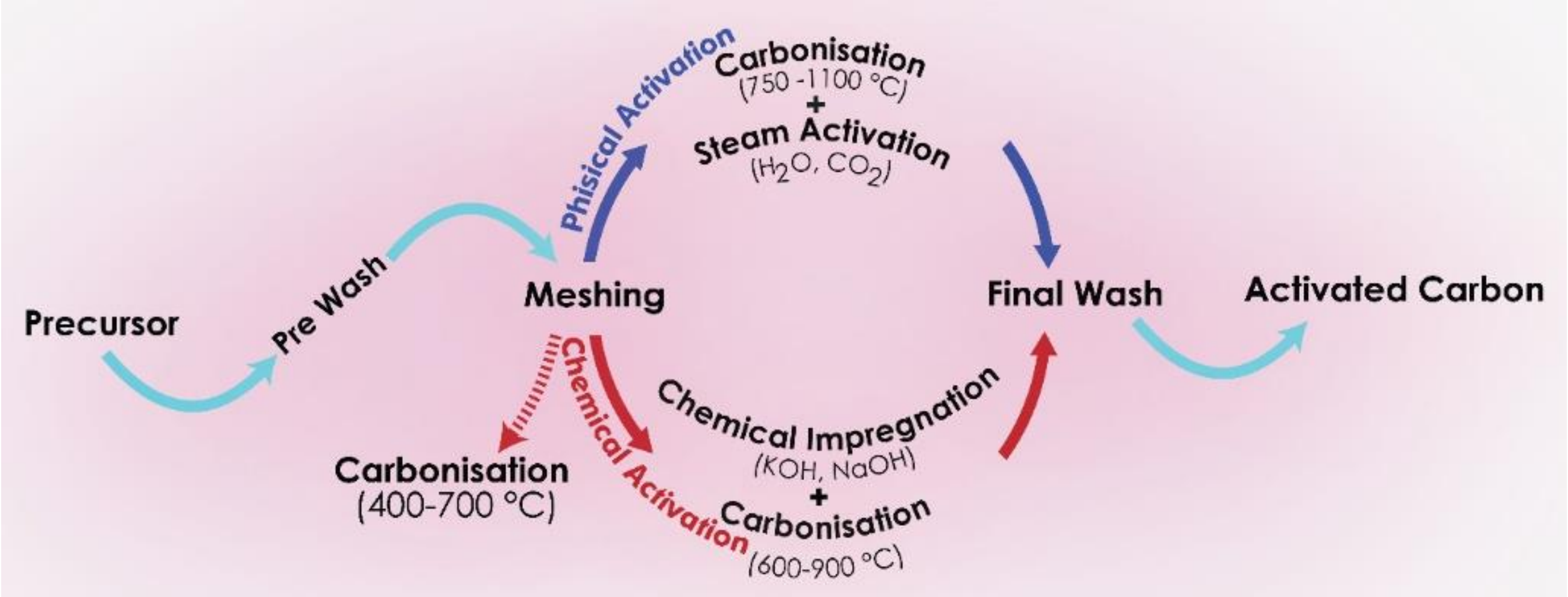

**Figure 5**. Schematic diagram introducing the preparation process of AC adsorbent

A high temperature of up to 650-700 °C is applied during carbonisation to evaporate and remove volatile components from the precursor under an oxygen-free environment. In other words, non-carbon ingredients are volatilized under inert gasses such as $N_2$ and Ar, which causes an increase in the carbon content of the carbonized samples [42]. In the next step, The carbonised product with a carbon content of higher than 85%, called char or biochar, will be activated to form fine solid cavities (i.e. porosity) inside their structure [55, 56]. Activation is involved with consecutive reactions inside the precursor's framework under the thermal condition [54]. The presence of some functional groups (e.g. alkyl species (=$CH_2$), C–C, C–O–C, and C–O–H), which can serve as active sites, is inevitably essential for the activation reactions to develop the mico/mesoporosity. The lack of these functional groups in the matrix structure of the carbonized precursors harms the formation of appropriate micropore distributions and consequently the gas storage capacity of the final AC [13, 23, 24, 30, 32, 36-42].

Here, we reviewed several conventional and novel production approaches for ACs comprising physical activation, chemical activation, combined phisico-chemical activation, hydrothermal carbonization, templating method together with practical techniques for assembly of ordered/structured adsorbents. The synthesis principles, type of carbon starting materials (precursors), pore structure formation, as well as the $CH_4$/$H_2$ storage and separation performance of the prepared ACs were analyzed under each preparation approach.

### 2.1 Physically activated carbons

Physical activation, also known as “thermal oxidation”, is considered an inexpensive approach for producing activated carbon without applying any chemical. This method can produce ACs with well-developed porosity in the solid structure favourable for gas adsorption requests. The primary purpose of such activation is to extend and open the incipient porosity developed during the carbonization [57]. This method mainly involves partial gasification of the carbon framework under controlled atmospheres of oxygen [5]/ air [5]/ carbon dioxide [58]/ steam [59] or their mixtures [60] at a high-temperatures of (700-1400 °C).

Among many preparation parameters, activation temperature and gas flow rate can effectively impress the final textural features of activated carbon [5, 54]. The activation temperature generally depends on the reactivity between the raw carbon materials and oxidant agents. However, it is recommended to hold the temperature at a medium point to assure a diffuse regime control of the reaction and prevent the fast gasification of the external surfaces [5]. In the case of gas flow rate, adjusted entering gas speed is required. With high gas flow rates, the physical agent does not deeply penetrate inside the precursor's structure, and the burn-off (i.e. amount of the carbons which react with oxygen and leave the solid structure as a CO or $CO_2$ gases) issue becomes more important than developing the porosity. On the other hand, a slow gas flow rate (below 10 ml/min) favours creating more uniform microporosity due to more even gasification of the whole carbon particle network [37]. The following reactions happen during the activation of the carbon materials under an oxidant environment of oxygen/ air (reaction 1), carbon dioxide (reaction 2) and steam (reaction 3), which result in the formation of highly porous structures [13]:

$$\mathbf{C + O_2 \rightarrow CO_2 \quad (1)}$$

$$\mathbf{C + CO_2 \rightarrow 2CO \quad (2)}$$

$$\mathbf{C + H_2O \rightarrow CO + H_2 \quad (3)}$$

Among many of the reported physically prepared ACs for gas storage requests [57, 61-63], spherical porous carbons (SPCs), shown in Figure 6, have been attracted great interest because of their high surface-to-volume ratio, good structural features and ample void space for encapsulating plenty of guest molecules [62, 64]. Generally, the SPCs are derived from renewable resources (e.g. starch, nutshells, glucose) via the $CO_2$ activation technique [62, 65-67]. Li *et al*. prepared starch-derived porous carbons with spherical morphology via an environmental-friendly and straightforward $CO_2$ activation without using any template [62]. The physical activation process of starch-based carbon spheres (CSs) was involved with two stages of (i) carbonisation, where the CSs were heated up to 500 °C in a tube furnace under flowing $N_2$ (60 ml. $min^{-1}$, 99.9% purity) for 180 min, afterwards (ii) the temperature raised to 1000 °C under $CO_2$ atmosphere (100 ml.$min^{-1}$, 99.9% purity) to activate the carbonized CSs for 150 min. Based on the SEM results depicted in Figure 6, the high-temperature-activated CSs were thermally and morphologically stable - the spherical morphology was maintained fixed with no significant changes after the high-temperature activation. They also reported an exceptionally high surface area of 3350 $m^2.g^{-1}$ and a high pore volume of 1.75 $cm^3.g^{-1}$ for the activated CS. Moreover, high uptakes values for $CH_4$ (16.7 $mmol.g^{-1}$, at 20 bar and 25 °C) and $H_2$ (6.4 wt.%, 22 bar and -196 °C) made these adsorbents a promising candidate for gas storage application.

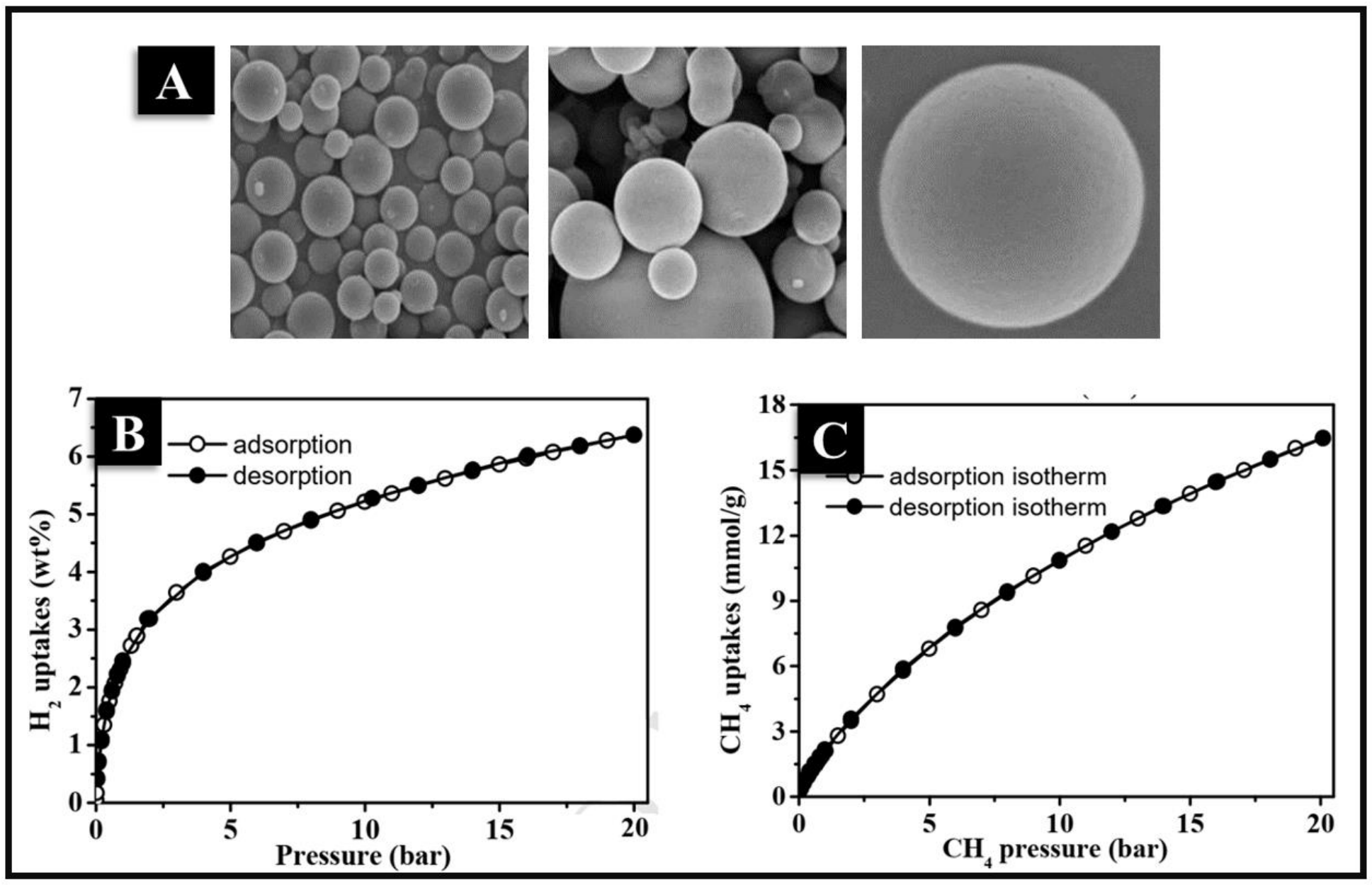

**Figure 6.** (A) SEM images for carbon spheres at different magnifications (from left to right 7000, 10000 and 40000 respectively), (B) Gravimetric $H_2$ adsorption isotherm at 30 °C, (C) gravimetric $CH_4$ adsorption reversibility at 30 °C. Reproduced from ref [62]. Copyright 2016, Elsevier Ltd.

The physical activation can be considered as an inexpensive green (chemical-free) approach to produce ACs with narrow pore size distribution and good physical stability. However, the gas storage capacity of the physically activated adsorbents, especially on a volumetric scale, is usually lower than those prepared with other methods [23, 68-70]. This might be related to the relatively low packing density of the ACs produced by this method. Therefore, a balance between the microporosity and balk/packing density should be kept for gas storage purposes, especially for onboard vehicles.

## 2.2 Chemically activated carbons

Chemical activation is believed to be one of the favourite approaches capable of developing low-cost and efficient micro/mesoporous carbonaceous adsorbents for gas storage requests. Chemical activation, which is also known as "wet oxidation", includes a series of chemical reactions between a solid precursor and a chemical activating agent during the heating process under an inert environment. The precursor + chemical agent complex is simultaneously thermalized and activated at the targeted temperature (400 to 800 °C) under nitrogen or argon atmosphere. The chemical activation is considered economical for preparing ACs due to the relatively lower activation temperature and shorter retained time than the physical activation. Moreover, the resultant ACs show to have a higher specific surface area, uniform pore size distribution (PSD), and relatively higher packing density; these advantages positively affect the volumetric adsorption capacity of the products. The last item is especially essential for onboard gas storage applications.

### 2.2.1 Chemical agent

A detailed study on the mechanism of different chemical activation reactions in the precursor + chemical agent complex can be found in Rodriguez et al. efforts [60, 71]. They reported that although all kinds of chemical agents react with the precursor, some apparent differences are observed at the end of the impregnation step in the resulting materials. The most common chemical agents can also be divided into three species; (i) the strong chemicals, including KOH and NaOH [57, 68, 72-74], (ii) the middle range activating agents include $K_2CO_3$ and $Na_2CO_3$ [75-77], and (iii) the weak alkaline activating agents, also known as salts of strong alkalis and weak acids, comprising $ZnCl_2$ and $H_3PO_4$ [57, 78-82]. KOH is the most effective activating chemical for synthesising ACs with an extremely high specific surface area among these alkaline activating agents. Potassium hydroxide can suppress the generation of tar, allow a lower activation temperature during the activating reaction, accelerate the removal of noncarbon components and enhance the reaction rate of pyrolysis [83]. Besides, reaction with KOH as a chemical agent widens the micropore, $ZnCl_2$ increases small mesoporosity and $H_3PO_4$ forms a more heterogeneous pore size distribution in the matrix structure of the final adsorbent [71].

The likely reactions that might be happening between KOH and carbon raw materials are presented in the following equations:

$$\mathbf{4KOH + {=}CH_2 \rightarrow K_2CO_3 + K_2O + 3H_2} \quad \mathbf{(4)}$$

$$\mathbf{K_2O + C \rightarrow 2K + CO} \quad \mathbf{(5)}$$

$$\mathbf{K_2CO_3 + 2C \rightarrow 2K + 3CO} \quad \mathbf{(6)}$$

The initial mixing of the precursor and the chemical agent, also called the “mass-transfer” step, is critical to reaching well-developed porous carbons for gas storage uses. The raw carbon materials can be mixed with the agent in an aqueous solution (impregnation method) or just directly by grinding in the solid phase (dry method). Depending on the application of interest, either of these mixing methods is efficient. As a general rule, activated carbon produced via the dry mixing of precursor + alkaline agents has a higher and more well-developed porosity than those made by the impregnation method [23, 68]. In a study by Byamba-Ochir et al. [55], they investigated the impact of dry and wet initial mixing of NaOH as the activating agent on the porous texture of anthracite-based carbons. It was shown that the activation of the dry mixed sample resulted in a higher BET surface area of 1375-2063 $m^2.g^{-1}$ than those prepared by the impregnation method (816-1763 $m^2.g^{-1}$). Moreover, since the dry process requires less time and energy consumption than the impregnation method, physical or dry mixing is preferable to produce activated carbons on an industrial scale.

Finding optimum synthesis conditions to achieve a high gas capacity is challenging since many different preparation parameters were involved in the chemical activation process. Moreover, a multivariate study including an empirical investigation and laboratory analysis is quite expensive and time-consuming. Practising soft computations and using their nonlinear mapping capability and lack of necessity for detailed mechanical knowledge seems to be an excellent choice to attain ACs with high stored gas capacity. Several well-known empirical models are used for similar material designs to maximize the corresponding adsorbent's gas storage characteristics [24, 82, 84, 85]. We investigated the effect of different preparation variables (activation agent ratio, activation temperature and retention time) on the specifications of the coal-based ACs. The $CH_4$ adsorption capacity of the ACs was examined experimentally, and the results were analyzed systematically through empirical models (see Figure 7) [24]. Mathematical optimization tools (i.e. response surface methodology and genetic algorithm) were used to find the best preparation condition where the highest gas adsorption capacity can be achieved.

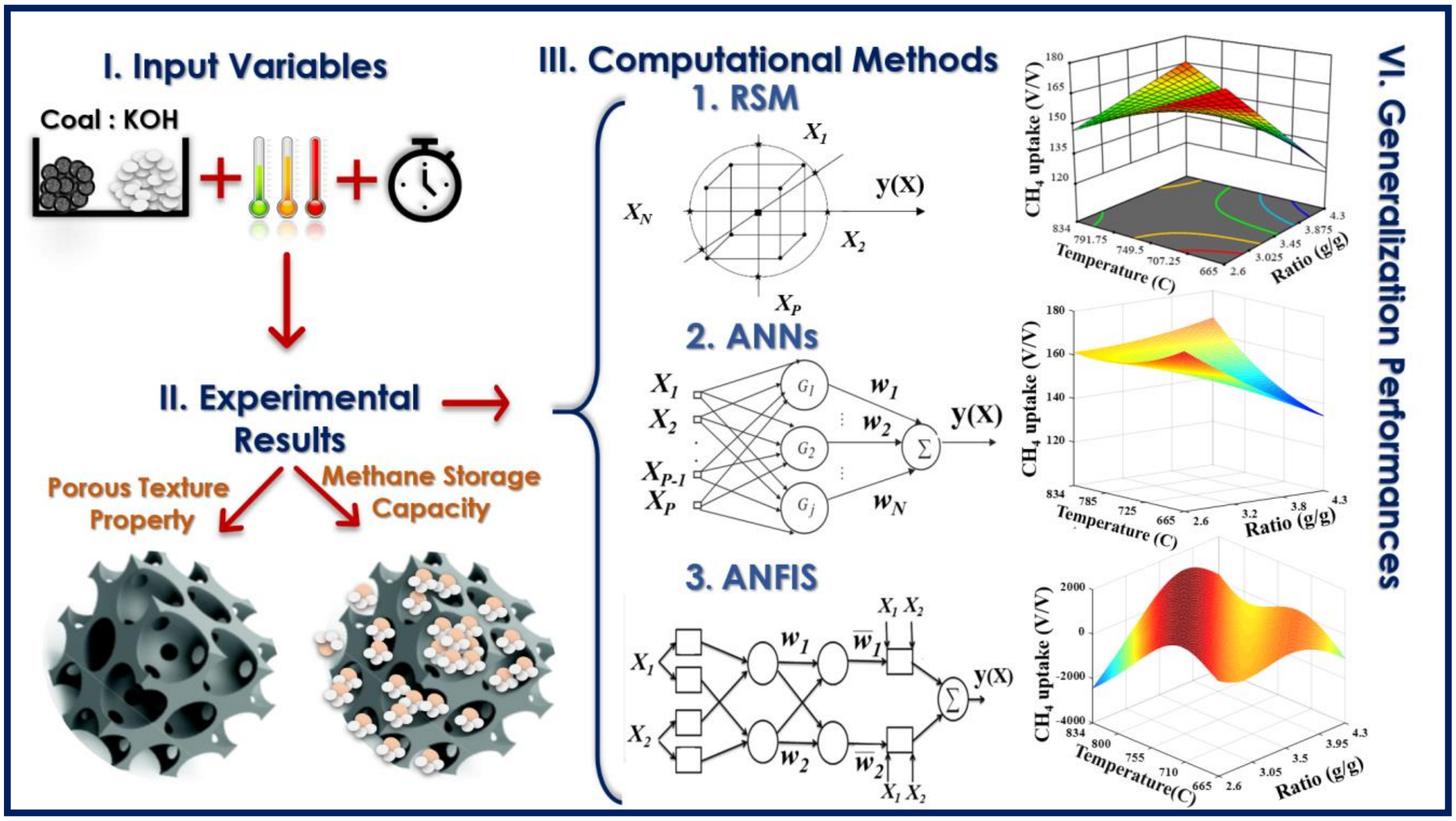

**Figure 7.** A schematic illustration of experimental and theoretical pathways. Reproduced from ref [24]. Copyright 2020, American Chemical Society.

In a similar study, Ahmadpour et al. applied the models of artificial neural networks and adaptive neuro-fuzzy interference systems to predict ACs porosity characterization (i.e. BET surface areas; $S_{BET}$, and micropore volumes; $V_{micro}$) [84]. They found that the models were able to predict the experimental data reasonably.

### 2.2.2 Raw materials

The pore structure and consequently gas adsorption capacity of ACs profoundly is affected by many parameters such as type of precursors, type and amount of chemical agents, carbonization temperature, and activation parameters (e.g. thermal condition and retention time) [69, 84]. Here, the effect of the preparation parameters of precursors and activation agents on the gas storage capacity of the prepared ACs are discussed.

Sawant *et al.* used raw and calcined petroleum cokes as a carbon precursor to prepare activated carbons on medium and pilot-scale by chemical activation method [45]. Results showed that raw petroleum coke was a favourable choice as a precursor, whereas the enhanced graphitic arrangement limits the applicability of the calcined petroleum coke. The prepared activated carbon exhibited a high specific urface area of 3578 $m^2.g^{-1}$ and high $CH_4$ storage capacity of 10.87 $mmol.g^{-1}$ at 30 °C and 37 bar, and $H_2$ capacity of 26.67 $mmol.g^{-1}$ at -196 °C and 3 bar. Based on the equilibrium adsorption data of $CO_2$, $CH_4$, CO, and $N_2$ measured on the prepared powder ACs at different temperatures, it was concluded that the adsorbents are promising applicants for gas separation. In another study [42], we prepared a series of ACs from carbonized coal tar pitch (CTP) using KOH as the activating agent to investigate the effect of two-stage treatment (acidification and carbonization) of the CTPs on the structural properties and $CH_4$ adsorption capacity of the prepared adsorbents. Depending on the thermal condition in the pre-carbonization step, the post-activated samples showed different $CH_4$ storage capacities. Among all ACs, a sample with $S_{BET}$ of 2261 $m^2.g^{-1}$ which carbonized at 600 °C for 2 h and later activated at 900 °C for 3 h, showed the maximum $CH_4$ uptake.

Past decades have witnessed a rapid development of synthetic polymers and related materials, including poly(vinylidene chloride)[86], phenyl-trimethylsilane [47], Polypyrrole [87], terephthalaldehyde [88], benzimidazole-linked polymers [48], polythiophene [89], porous coordination polymers [90], as potential precursors for the preparation of ACs for gas storage and purification applications. Cai et al. synthesized poly (vinylidene chloride)-based carbon (PC) with ultrahigh microporosity by simple carbonization and moderate KOH activation [86]. The obtained samples possessed a high surface area (up to 2150 $m^2.g^{-1}$) and pore volume (up to 0.9 $cm^3.g^{-1}$). The $CH_4$ uptake value for the activated PC was reported up to 10.25 $mmol.g^{-1}$ (16.4 wt.% or 147

V.V$^{-1}$) at 25 °C and 20 bar. Moreover, they measured a large $H_2$ adsorption capacity of 4.8 wt.% at -196 °C and 20 bar for the activated PC. The sample was shown to have a moderate selectivity for $CO_2/CH_4$ and $CH_4/N_2$. Attia and colleagues [87] developed a flexible AC by chemical activation of commercial viscose rayon cloth fibre. The final adsorbent showed a high surface area of 2000 m$^2$.g$^{-1}$ and superior storage capacity for $CH_4$ (7.5 mmol.g$^{-1}$, at 20 bar and 25 °C) and $H_2$ (4 wt.%, at 20 bar and -196 °C). They also reported a high separation selectivity of 15.9 for $CO_2/CH_4$ based on the ideal adsorbed solution theory (IAST) model.

Shirazani and co-worker's synthesized nanoporous carbon spheres by KOH activation of starch as the starting carbon source to reach the high $CH_4$ uptake [49]. The best nanoporous sample with a carbon: KOH ratio of 1:4 showed a high specific surface area of 2222.3 m$^2$.g$^{-1}$, pore volume up to 1.72 cm$^3$.g$^{-1}$ and $CH_4$ storage capacity of 12.8 mmol.g$^{-1}$ (at 25 °C and 35 bar). In another study, Jung *et al*. prepared a set of porous carbons by KOH activation of fruit by-products (i.e. Mandarin peels) [52]. A scalable synthesis approach was designed for recycling and converting these economically bio-wastes to valuable products, wherein the impact of carbonization and activation temperatures and activating agents was thoroughly investigated. As-made nanoporous carbons demonstrated good textural properties such as a high surface area of ~2500 m$^2$.g$^{-1}$ and pore volume of 1.04 cm$^3$.g$^{-1}$. Experimental measurements proved that these nanoporous materials possessed a promising sorption capacity for $H_2$ (6.1 wt.% at -196 °C and 25 bar) and $CH_4$ (9.65 mmol.g$^{-1}$ at 25 °C and 25 bar) together with an excellent gas separation performance for binary mixtures of ($CO_2$ + $CH_4$), ($CH_4$ + $N_2$) and $CO_2$ + $N_2$).

An almost new term of “Physio-chemical” methodology generally represents a combination of two conventional physical and chemical activation approaches, involving chemical impregnation of carbon precursors with an activating agent, followed by a physical activation step under an oxidizing gas atmosphere (mainly $CO_2$). Well-regulated porous textural properties and surface modification ability are considered the advantages of this mixed-method [91]. In this method, the chemical agent generates small micropores while the physical activation makes the pores' size homogenus and narrows pore size distribution around an optimum diameter/width. Up to date, several valuable studies have been reported for the synthesis of physio-chemically activated carbons for gas storage/separation uses [91-93]. Preacher and coworkers prepared several ACs in powder, granule, and monolith form using a both chemical and phisical activation approaches to improve the gas adsorption capacity of solids [91]. They used $H_3PO_4$ and $ZnCL_2$ as the chemical agents, and $CO_2$ was used as a physical activator. The $H_3PO_4$ and $ZnCL_2$-activated samples respectively showed acceptable $CH_4$ adsorption values of 155 $V.V^{-1}$ and 165 $V.V^{-1}$ (25 °C and 35 bar). Based on the experimental observations, the maximum gas storage values were verified for the series prepared with $ZnCl_2$ rather than the prepared sample using $H_3PO_4$. These results were justified by the formation of narrower pore width distribution centered around 0.8 nm by $ZnCl_2$ activation, such PSD is about twice the dimension of the methane molecule and provides a fine space for optimizing the methane storage capacity [93].

Although the combined activation approach enjoys the advantages of both originated physical and chemical activation methods, it is a complex and energy-consuming technique. As a result, there is a considerable gap between laboratory studies and large-scale production for industry application [83].

**Table 2.** Non-renewable porous carbons prepared by activation for gas storage.

| Precursor + Activator | Porosity | | | $\rho$ packed ($g.cm^{-3}$) | $H_2$ uptake (wt.%) | $CH_4$ Storage Capacity ($V.V^{-1}$) | Selective Separation Capacity (SSC) | | Measurement Condition | Reference |
|---|---|---|---|---|---|---|---|---|---|---|
| | $S_{BET}$ ($m^2.g^{-1}$) | Vt / $V_p$ ($cm^3.g^{-1}$) | $d_{ave}$ (nm) | | | | $CO_2/CH_4$ | $CH_4/N_2$ | | |
| DOE targets | - | - | | - | 5.5 | 350 | - | - | $H_2$ @ -196 °C, 20 bar<br>$CH_4$ @ 25 °C, 35 bar | [94, 95] |
| 1,3 bis(cynomethyl imidazolium) chloride + KOH | 1317 | 0.64 / 0.43 | 0.59 | - | 2.96 | - | - | - | $H_2$ @ -196 °C, 1 bar | Sethia et al.[96] (2016) |
| Poly (vinylidene chloride) (PVDC) + KOH | 2151 | 0.90 / 0.85 | 1.44 | 0.64 | 5 | 147 | - | - | $H_2$ @ -196 °C, 20 bar<br>$CH_4$ @ 25 °C, 20 bar | Cai et al.[86] (2014) |
| Polypyrrole + $ZnCl_2/NH_4Cl$ | 1911 | 0.85 / 0.60 | 0.85 | - | 4 | 7.5 mmol. $g^{-1}$ | 3.4 | - | $H_2$ @ -196 °C, 20 bar<br>$CH_4$ @ 25 °C, 20 bar<br>SSC @ 25 °C, 20 bar | Attia et al.[87] (2020) |
| Ethylene tar + KOH | 2465 | 1.12 / 0.89 | 0.80 | 0.60 | - | 150 | - | - | $CH_4$ @ 25 °C, 35 bar | Casco et al.[32] (2015) |
| Decanted oil + KOH | 2700 | 1.23 / 0.93 | 0.80 | 0.62 | - | 160 | - | - | $CH_4$ @ 25 °C, 35 bar | Casco et al.[32] (2015) |
| Mesophase pitches + KOH | 3290 | 2.25 / 1.10 | - | 0.53 | - | 156 | - | - | $CH_4$ @ 25 °C, 35 bar | Casco et al.[39] (2015) |
| Benzimidazole-linked polymers (BILP) + KOH | 1630 | 0.66 / 0.59 | 0.70 | - | - | 1.7 mmol. $g^{-1}$ | 13 | - | $CH_4$ @ 25 °C, 1 bar<br>SSC @ 0 °C , 1 bar | Ashourirad et al.[48] (2015) |
| Nitrogen-rich BILP + KOH | 2059 | 0.89/ 0.74 | 0.90 | - | - | 1.5 mmol. $g^{-1}$ | 10 | - | $CH_4$ @ 25 °C, 1 bar<br>SSC @ 0 °C , 1 bar | Ashourirad et al.[48] (2015) |
| Petroleum tar pitch and powdered coal + KOH | 1044 | 0.5 / 0.46 | 0.98 | 0.42 | - | 5.05 mmol. $g^{-1}$ | - | 2.23 | $CH_4$ @ 25 °C, 40 bar<br>SSC @ 25 °C, 1 bar | Arami-Niya et al.[46] (2016) |
| Nitrogen-rich polymer + KOH | 2146 | 1.96 / 0.91 | 3.20 | - | - | 2.12 mmol. $g^{-1}$ | 47.1 ($C_2H_2/CH_4$) | | $CH_4$ @ 0°C , 1.13 bar<br>SSC @ 25 °C, 1.13 bar | Wang et al.[88] (2016) |
| Hyper crosslinked porous polymers + KOH | 3101 | 184 / 1.40 | 1.40 | - | 3.25 | 2.76 mmol. $g^{-1}$ | - | - | $H_2$ @ -196°C , 1.13 bar<br>$CH_4$ @ 0 °C , 1.13 bar | Zhang et al.[47] (2017) |
| Calcined petroleum cokes + KOH | 3575 | 191 / 1.26 | 2.10 | - | - | 1.96 mmol. $g^{-1}$ | 2.1 | 2.8 | $CH_4$ @ 24°C , 1.13 bar<br>SSC @ 15 °C, 1.13 bar | Sawant et al.[45] (2017) |
| Carbon black + KOH | 1639 | 1.25 / 0.66 | 0.82 | - | 234.71 $cm^3.g^{-1}$ | - | - | - | $H_2$ @ -196 °C, 1 bar | Heo et al.[3] (2019) |
| Anthracite + KOH | 2160 | 0.75 | 0.98 | 0.53 | - | 175 | - | - | $CH_4$ @ 25 °C, 40 bar | Mirzaei et al.[24] (2020) |
| **Coal tar pitch + KOH** | **2261** | **0.74** | **2.17** | **0.70** | - | **184** | **-** | **-** | **$CH_4$ @ 25 °C, 40 bar** | **Mirzaei et al.[42] (2020)** |
| **Anthracite + KOH/urea treatment** | **1965** | **0.56** | **2.07** | **0.63** | **0.26 mmol. $g^{-1}$** | **190** | **-** | **-** | **$H_2$ @ 25 °C, 20 bar<br>$CH_4$ @ 25 °C, 40 bar** | **Mirzaei et al.[25] (2020)** |
| **Viscose rayon fabric + $ZnCl_2$/$NH_4Cl$ + $CO_2$** | **1205** | **- / 0.46** | **0.78** | - | **1.68** | **0.72 mmol. $g^{-1}$** | **4.5** | **-** | **$H_2$ @ -196 °C, 1 bar<br>$CH_4$ @ 25 °C, 1 bar<br>SSC @ 25 °C, 1 bar** | **Kostoglou et al.[92] (2017)** |
| **Polypyrrole + KOH** | **3225** | **2.43 / -** | **1.3** | - | **6.5** | **-** | **-** | **-** | **$H_2$ @ -196 °C, 20 bar** | **Adeniran et al.[97] (2015)** |

**Table 3.** Renewable porous carbons prepared by activation for gas storage.

| Precursor + Activator | Porosity | | | $\rho$ packed ($g.cm^{-3}$) | $H_2$ Uptake (wt.%) | $CH_4$ Storage Capacity ($V.V^{-1}$) | Selective Separation Capacity (SSC) | | Measurement Condition | Reference |
|---|---|---|---|---|---|---|---|---|---|---|
| | $S_{BET}$ ($m^2.g^{-1}$) | Vt / $V_p$ ($cm^3.g^{-1}$) | $d_{ave}$ (nm) | | | | $CO_2/CH_4$ | $CH_4/N_2$ | | |
| Oil-tea seed shell + $NaNH_2$ | 774 | 0.32 / 0.27 | 0.54 | - | - | 0.94 mmol. $g^{-1}$ | 12.48 | 4.9 | $CH_4$ @ 25 °C, 1 bar<br>SSC @ 25 °C, 1 bar | Zhang et al.[98] (2019) |
| Starch + KOH | 2222 | 1.73 / - | 3.08 | - | - | 12.81 mmol.$g^{-1}$ | - | - | $CH_4$ @ 25 °C, 35 bar | Shirazani et al.[49] (2020) |
| Date seed + KOH | 2609 | - / 0.70 | 0.63 | 0.69 | - | 13.2 mmol. $g^{-1}$ | - | - | $CH_4$ @ 25 °C, 35 bar | Altwala et al.[51] (2020) |
| Dry algae + KOH | 1247.2 | 0.69 / - | 0.51 | - | - | 1.3 mmol. $g^{-1}$ | 10.1 | 4.1 | $CH_4$ @ 25 °C, 1 bar<br>SSC @ 25 °C, 1 bar | Wang et al.[53] (2018) |
| Chitosan + KOH | 2513 | 1.08 / 0.97 | 1.25 | - | 2.88 | - | - | - | $H_2$ @ -196 °C, 1 bar | Wrobel et al.[50] (2015) |
| Mandarin peels + KOH | 2426 | 1.04 / 0.96 | 0.78 | - | 6.10 | 6.07 mmol. $g^{-1}$ | 13.6 | - | $H_2$ @ -196 °C, 25 bar<br>$CH_4$ @ 25 °C, 25 bar<br>SSC @ 25 °C, 25 bar | Jung et al.[52] (2020) |
| Coconut shells + KOH | 1932 | 1.39 / 0.85 | 0.17 | - | 11.16 | - | - | - | $H_2$ @ -196 °C, 60 bar | Park et al.[99] (2019) |
| Neolamarckia cadamba + KOH | 3246 | 1.33 / 1.08 | 1.59 | - | 2.81 | - | - | - | $H_2$ @ -196 °C, 1 bar | Hu et al.[100] (2019) |
| Starch + $CO_2$ | 3350 | 1.75 / 1.67 | 1.19 | - | 6.4 | 10.7 mmol. $g^{-1}$ | - | - | $H_2$ @ -196 °C, 20 bar<br>$CH_4$ @ 25 °C, 20 bar | Li et ai. [62] (2016) |
| Agricultural waste + $N_2$ | 790 | 0.32 / 0.25 | 0.70 | - | 1.29 | 1.38 mmol. $g^{-1}$ | 15 | - | $H_2$ @ -196 °C, 20 bar<br>$CH_4$ @ 40 °C, 1 bar<br>SSC @ 40 °C, 1 bar | Park et al.[61] (2018) |
| Crab shell + $CO_2$ | 1708 | - / 0.89 | 0.55 | - | - | 1.74 mmol. $g^{-1}$ | - | 6.8 | $CH_4$ @ 25 °C, 1 bar<br>SSC @ 25 °C, 1 bar | Kim et al.[58] (2018) |
| Phenolic resin + $CO_2$ | 1107 | 0.46 / 0.39 | 0.70 | - | 2.48 | - | - | - | $H_2$ @ -196 °C, 20 bar | Tian et al.[101] (2020) |
| Poly (furfuryl alcohol) + $CO_2$ | 1530 | 0.67 / 0.42 | 3.6 | - | 1 mmol. $g^{-1}$ | - | - | - | $H_2$ @ 25 °C, 20 bar | He et al.[102] (2014) |
| Coconut shells + $ZnCl_2$ + $CO_2$ | 1599 | 0.77 / 0.70 | - | 0.44 | - | 199 $cm^3.g^{-1}$ | - | - | $CH_4$ @ 25 °C, 35 bar | Prauchner et al.[91] (2016) |
| Coconut shells + $H_3PO_4$ + $CO_2$ | 2191 | 1.12 / 0.78 | - | 0.36 | - | 220 $cm^3.g^{-1}$ | - | - | $CH_4$ @ 25 °C, 35 bar | Prauchner et al.[91] (2016) |

### 2.3 Surface-modified activated carbons

Besides the direct effect of carbon's microporosity on gas storage capacity, modifying the solid's surface chemistry (surface functionalization) can improve the adsorbent-adsorbate interaction and gas adsorption capacity. Nitrogen (N), and fluorine (F) containing contents, as well as other cations, are some of the functional groups which can be introduced (doped) into the carbon skeleton to create some momentary polarisation in the gas molecules [25, 103-107]. However, the impact of these heteroatom-dopings on the storage efficiency of gases, especially $CH_4$ and $H_2$, is rarely studied and is often controversial. Herein, we briefly summarised the recent investigations to clarify the influences of surface chemistry on the gas uptake of carbons.

Nitrogen/oxygen doping is a common modification method to introduce basic sites into the carbon framework and enhance the polarity of the carbon surface [25, 106]. The high electronegativity of nitrogen/oxygen-containing surface groups might affect the non-polar gas molecules near carbon pores, which results in a higher uptake of $H_2$ and especially $CH_4$ molecules [25, 104]. The regular tetrahedron structure of a $CH_4$ molecule, the central C atom exhibits high electronegativity restricted by four electropositive H atoms [38]. The electronegative N and O atoms of the surface functional groups contribute their electrons to the H atoms of $CH_4$ and promote the adsorption of $CH_4$. On the other hand, a repulsive force is evident between the C atom of the $CH_4$ molecule and the electronegative atoms of the functional groups, resulting in a reduction in the amount of $CH_4$ uptake [38]. Therefore, the introduction of functional groups on the surface of ACs has a two-fold impact (attractive and repulsive interactions) on the $CH_4$ adsorption characteristics of the corresponding adsorbents (Figure 8) [38, 108].

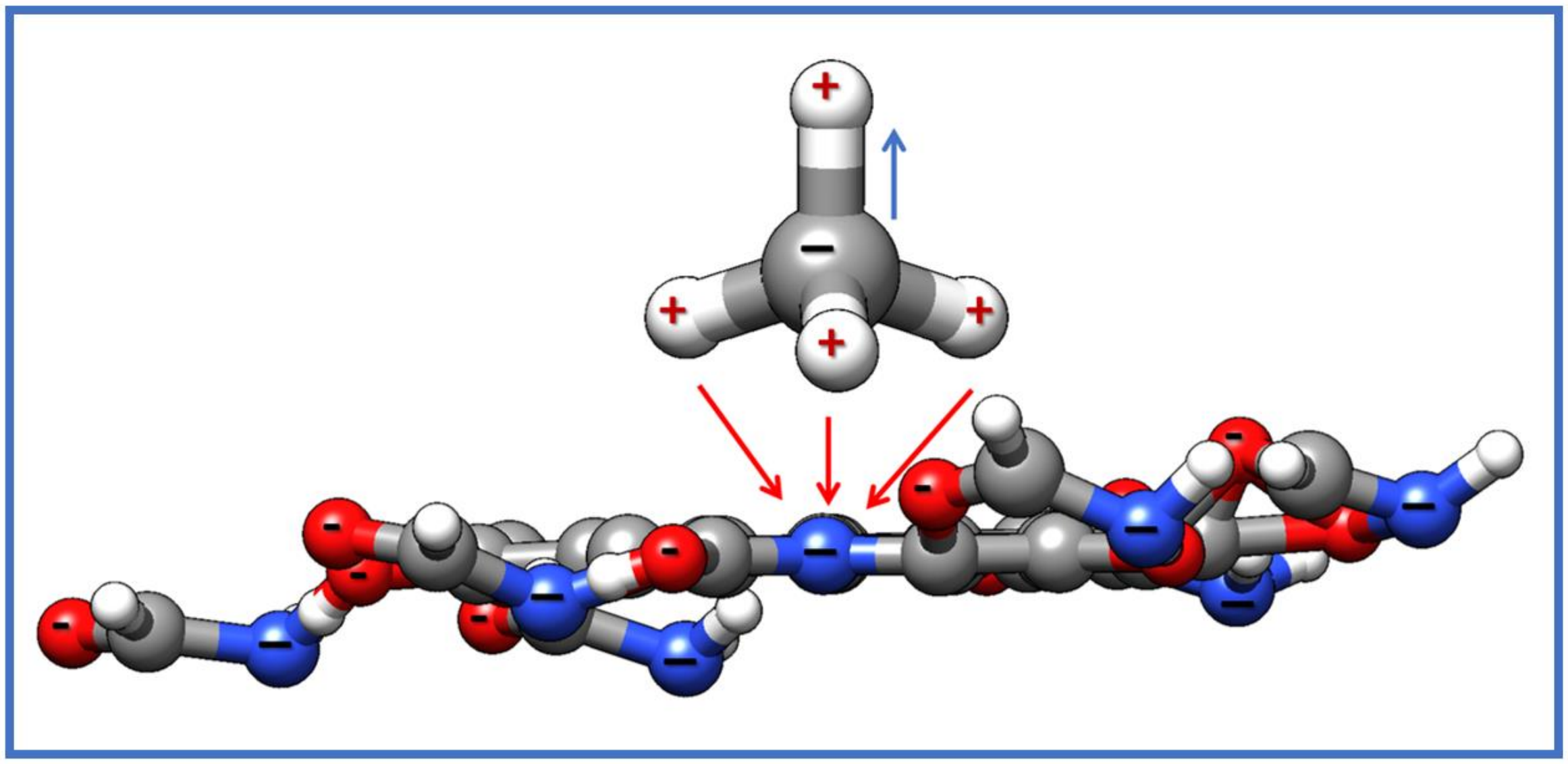

**Figure 8.** Schematic display of the interaction between the adsorbate ($CH_4$) and charged atoms in the functional group of the AC-OX-U adsorbent. The visualized colors of gray, white, red and blue represent the atoms of carbon, hydrogen, oxygen, and nitrogen, respectively.

Introducing different functional groups on the surface of the ACs may have either electron-withdrawing or -donating effects. This issue influences the $H_2$ adsorption capacity of adsorbents [109]. In other words, $H_2$ has a weak bipole, so the adsorption of this molecule toward the solid structure of the modified ACs is profoundly affected by the electronegativity of the surface functional groups. The N- (3.04) and O-containing (3.44) functional groups exhibit strong electronegativity, but C atoms (2.55) in the matrix structure of ACs show an ultra-strong electropositivity [38, 110]. Therefore, the nitrogen and oxygen-containing group introduce negative-charge characteristics on the surface, while the carbon exhibits relatively positive-charged characteristics. As a result, this electronegativity gap will electrically attract the $H_2$ molecule into the modified carbonaceous adsorbents [110].

We have previously studied the effect of oxygen/urea (as a source of nitrogen-containing components) on the gas storage capacity of porous carbons [25]. After functionalization of the carbon surface, the volumetric $CH_4$ adsorption capacity of the corresponding sample was increased by 11 % to 190 $cm^3.cm^{-3}$ at 25 °C and 40 bar. However, it was concluded that larger micropore volume is preferable to increase the $H_2$ storage capacity of ACs, which shows the inductive effect of surface chemistry as a secondary role in $H_2$ uptake. Similar results were reported by other authors [104, 111]. Li et al. developed cost-efficient ACs for gas storage and separation purposes via KOH activation of anthracite followed by urea treatment as a surface modification step [106]. The as-prepared carbon samples maintained developed microporosity of 0.20-0.32 $cm^3.g^{-1}$ and a good BET surface area of 912-1615 $m^2.g^{-1}$. The gas adsorption measurements showed that although nitrogen doping contributes significantly to the selectivity of $CO_2/CH_4$, 8.09 based on IAST calculation, it almost harms the $CH_4$ uptake of the N-dope carbons (1.46 $mmol.g^{-1}$ at 25 °C and 1 bar) compared to those simple counterpart samples (1.59 $mmol.g^{-1}$ at 25 °C and 1 bar). This observation might be interpreted by the fact that methane molecule possesses relatively low polarizability and heteroatom doping does not facilitate $CH_4$ storing [106].

Blankenship II et al. experimentally optimized the fabrication of ACs to generate a set of porous carbons that simultaneously enjoyed a high surface area and high level of microporosity along with a high oxygen-rich nature [104]. They observed a positive effect of having extra oxygen functional groups at low pressure, where interaction between the hydrogen and surface becomes more important than at higher pressure where hydrogen uptake is more likely to occur via space-filling mechanisms. Such AC samples exhibited enhanced gravimetric hydrogen storage capacity of up to 8.9 wt% (total uptake) and 7.2 wt.% (excess uptake) at -196 °C and 30 bar. The result of hydrogen storage value is among the highest ever reported for porous carbons; the best previous reports have been in the range of 4–7.5 wt% [21, 112-114]. Table 4 summarizes some of the best activated carbons modified with functional groups or heteroatoms.

**Table 4.** Surface-modified porous carbons prepared for gas storage.

| Precursor + Activator | Porosity | | Functional groups/ Heteroatom | $H_2$ Uptake (wt.%) | $CH_4$ Storage Capacity ($V.V^{-1}$) | Measurement Condition | Reference |
|---|---|---|---|---|---|---|---|
| | $S_{BET}$ ($m^2.g^{-1}$) | Vt / $d_{ave}$ ($cm^3.g^{-1}$/ nm) | | | | | |
| Activated carbon fibers | 1053 | 0.50 / - | Fluorine | 1.32 | - | $H_2$ @ 30 °C, 100 bar | Lee et al. [115] (2007) |
| Activated carbon fibers | 1468 | 0.70 / - | Ni | 1.63 | - | $H_2$ @ 30°C, 100 bar | Lee et al. [115] (2007) |
| Poly Acrylonitrile + KOH | 2500 | 0.75 / 1.60 | Fluorine | - | 150 | $CH_4$ @ 25 °C, 35 bar | Im et al.[116] (2009) |
| Anthracite + KOH | 1320 | 0.62 / - | Urea ($CO(NH_2)_2$) | - | 1.5 $mmol.g^{-1}$ | $CH_4$ @ 25 °C, 1 bar | Li et al.[106] (2013) |
| Nanoporous carbons* | 3184 | 1.43 / - | -OH | - | 10.30 $mmol.g^{-1}$ | $CH_4$ @ 25 °C, 60 bar | Lu et al. [38] (2014) |
| Nanoporous carbons* | 3598 | 1.35 / - | $-NH_2$ | - | 9.85 $mmol.g^{-1}$ | $CH_4$ @ 25 °C, 60 bar | Lu et al. [38] (2014) |
| Nanoporous carbons* | 3548 | 1.36 / - | -COOH | - | 9.50 $mmol.g^{-1}$ | $CH_4$ @ 25 °C, 60 bar | Lu et al. [38] (2014) |
| Hydrochar + KOH | 3771 | 1.75 / 1.20 | O | 8.90 | - | $H_2$ @ -196 °C, 30 bar | Scott et al. [104](2017) |
| Glucose + $NaHCO_3$ | 1002 | 0.90 / 3.16 | MgO nanoparticles | - | 12 $mmol.g^{-1}$ | $CH_4$ @ 25 °C, 30 bar | Ghosh et al. [107] (2018) |
| Commercial AC | 825 | 0.24 / - | Nitric acid + Ni | 0.26 $mmol.g^{-1}$ | 3.14 $mmol.g^{-1}$ | $H_2$ @ 30 °C, 10 bar<br>$CH_4$ @ 30 °C, 9 bar | Rezvani et al. [117](2019) |
| Waste Wool + KOH | 862 | 0.50 / - | N | - | 1.70 $mmol.g^{-1}$ | $CH_4$ @ 0 °C, 1 bar | Li et al.[118] (2019) |
| Anthracite + KOH | 1878 | 1.14 / 1.91 | -OH | 0.44 $mmol.g^{-1}$ | 124.39 | $H_2$ @ 25 °C, 20 bar<br>$CH_4$ @ 25 °C, 40 bar | Mirzaei et al. [25] (2020) |
| Anthracite + KOH | 1924 | 0.92 / 1.89 | $-NH_2$ | 0.39 $mmol.g^{-1}$ | 178.75 | $H_2$ @ 25 °C, 20 bar<br>$CH_4$ @ 25 °C, 40 bar | Mirzaei et al. [25] (2020) |
| Anthracite + KOH | 1695 | 0.88 / 2.07 | -C(O)NH | 0.26 $mmol.g^{-1}$ | 190 | $H_2$ @ 25 °C, 20 bar<br>$CH_4$ @ 25 °C, 40 bar | Mirzaei et al. [25] (2020) |
| Coal + KOH | 3037 | 1.30 / - | Ammonia + Boron | 4.14 | - | $H_2$ @ -196 °C, 1 bar | Kopac et al. [119](2020) |
| Lignite + KOH | 500 | 0.046 / - | N | 0.50 | - | $H_2$ @ 30 °C, 30 bar | Han et al. [120](2021) |

*** This symbol represents theoretical study.**

## 2.4 Hydrothermal carbons

Hydrothermal Carbonization (HTC) is a well-established thermochemical synthesis approach for making functional carbon materials with high surface area and controlled porosity from wet or dry carbohydrates or lignocellulosic biomass [121]. During hydrothermal carbonization, water media acts as both solvent and catalyst, helping hydrolysis reactions and decomposition of the biomass into smaller fragments, which results in the formation of coal-like solid materials. Over the past decade, HTC considered a powerful technique for synthesising carbons from biomass [121, 122].

Preparation parameters such as the nature of the precursor and its thermal stability, temperature, residence time, and operation pressure should be optimised to reach the maximum efficiency in the HTC method [123]. As a general rule, increasing the temperature and pressure of the hydrothermal conditions increases the conversion of the biomass feedstock into carbon particles [76].

One of the main restrictions associated with the HTC technique is the relatively poor textural characteristics of the products (i.e. low surface area and undeveloped porosity). Thus the prepared adsorbents can not efficiently be applied for gas storage and separation. Post-activation of the hydrochars might be considered as a solution for producing ACs with high surface area [124]. Yuan et al. synthesized a series of nitrogen-doped porous carbons from a mixture of algae and glucose via hydrothermal carbonization and post-chemical activation [122]. They observed that post-KOH activation leads to a significant increase in the BET surface area of the ACs from 284 to 1538 $m^2.g^{-1}$. It was concluded that the products' $CH_4$ (2.5 $mmol.g^{-1}$ at 0 °C and 1 bar) and $H_2$ (0.1 $mmol.g^{-1}$ at 0 °C and 1 bar) storage capacity were primarily governed by narrow micropore volume and BET surface area. This study further supports the standpoint that post-synthesis activation is an efficient way to develop the porosity of carbon frameworks.

The HTC's temperature strongly influences the porosity of the post-activated carbons [124]. When the HTC is conducted in the temperature range of 200 to 250 °C, two factors of a higher density of carbonyl components and the low degree of aromatization facilitate the development of porosity during KOH activation. Falco and colleagues evaluated the impact of different HTC temperatures on porosity and gas adsorption properties of biomass-derived ACs [121]. They prepared HTC carbons, modified with an additional KOH activation step, from a pure monosaccharide (i.e. glucose), its polymer (i.e. cellulose) and lignocellulosic biomass (i.e. rye straw) under different thermal conditions. Results showed the HTC temperature is a highly influential parameter affecting the development of the ACs porosity and PSDs. It was reported that the samples prepared at medium HTC temperature (180–240°C) possessed higher total/micropore volume (in the range of 0.94–1.21 $cm^3.g^{-1}$), whilst those obtained under higher HTC temperature (upper that 280 °C) exhibited lower microporosity development (lower than 0.95 $cm^3.g^{-1}$) after the KOH activation.

Blankenship and Mokaya studied the preparation of HTC porous carbons from cigarette filters/butts for $H_2$ storage [125]. They showed that ACs produced via sequential processes of HTC and KOH-activation owned an ultra-high surface area of 4300 $m^2.g^{-1}$ and pore volume of 2.09 $cm^3.g^{-1}$ arising almost entirely (49%) from micropores. Due to the combined effects of super-high microporosity and an oxygen-rich nature of the ACs surface, the adsorbent exhibited unprecedentedly high $H_2$ storage capacity of 9.4 wt.% at -196 °C and 20 bar, rising to total uptake of 11.2 wt.% at 40 bar.

According to what has been reviewed in this section, hydrothermal carbonization is almost a new-conversion technique, ultimately converting waste materials to value-added products for practical uses, especially in the gas storage field. Table 5 summarizes some of the best samples fabricated by the HTC method.

**Table 5.** Porous carbons prepared by the HTC method for gas storage.

| Precursor + Activator | Porosity | | | ρ packed ($g.cm^{-3}$) | $H_2$ Uptake (wt.%) | $CH_4$ Storage Capacity ($V.V^{-1}$) | Measurement Condition | Reference |
|---|---|---|---|---|---|---|---|---|
| | $S_{BET}$ ($m^2.g^{-1}$) | $V_t$ / $V_p$ ($cm^3.g^{-1}$) | $d_{ave}$ (nm) | | | | | |
| D-glucose + KOH | 2210 | 1.21 / 0.90 | 1.22 | 0.38 | - | 86 | $CH_4$ @ 25 °C, 40 bar | Falco et al.[121] (2013) |
| Cellulose + KOH | 2250 | 1.26 / 0.90 | 0.55 | 0.39 | - | 86 | $CH_4$ @ 25 °C, 40 bar | Falco et al.[121] (2013) |
| Rye straw + KOH | 2200 | 1.11 / 0.92 | 0.52 | 0.43 | - | 105 | $CH_4$ @ 25 °C, 40 bar | Falco et al.[121] (2013) |
| Sucrose + $K_2CO_3$ | 1375 | 0.63 / 0.58 | 0.70 | 0.76 | - | 90 | $CH_4$ @ 25 °C, 10 bar | Mestre et al.[76] (2014) |
| Sucrose + KOH | 2431 | 1.14 / 0.90 | 0.77 | 0.36 | - | 40 | $CH_4$ @ 25 °C, 10 bar | Mestre et al.[76] (2014) |
| Mixture of algae/ glucose + KOH | 1192 | 0.54 / 0.48 | 1.2 | - | 0.08 $mmol.g^{-1}$ | 2.68 $mmol.g^{-1}$ | $H_2$ @ -196 °C, 1 bar<br>$CH_4$ @ 0°C, 1 bar | Yuan et al. [122] (2016) |
| Mixture of algae/ glucose + KOH | 1534 | 0.69 / 0.57 | 1.41 | - | 0.11 $mmol.g^{-1}$ | 2.15 $mmol.g^{-1}$ | $H_2$ @ -196 °C, 1 bar<br>$CH_4$ @ 0 °C, 10 bar | Yuan et al. [122] (2016) |
| Fresh cigarette Filter + KOH | 4113 | 1.87 / 0.79 | 0.60 | - | 3 | - | $H_2$ @ -196 °C, 1 bar | Blankenship et al.[125] (2017) |
| Smoked cigarette filters+ KOH | 4310 | 2.09 / 1.71 | 0.64 | - | 4 | - | $H_2$ @ -196 °C, 1 bar | Blankenship et al.[125] (2017) |
| Sucrose + KOH | 2240 | - / 1 | 0.65 | - | 0.18 | - | $H_2$ @ 25 °C, 20 bar | Schaefer et al.[126] (2016) |
| Wood sawdust + KOH | 2797 | 1.74 / - | 0.80 | - | 6.1 | - | $H_2$ @ -196°C, 20 bar | Adeniran et al.[97] (2015) |
| Potato starch + KOH | 3220 | 2.27 / 1.21 | 0.85 | - | 6.4 | - | $H_2$ @ -196°C, 20 bar | Sevilla et al.[127] (2016) |
| Mixtures of polypyrrole/raw sawdust + KOH | 3815 | 2.18 / 0.66 | 1.2 | - | 12.6 | - | $H_2$ @ -196°C,100 bar | Balahmar et al. [128](2019) |
| Lignin waste + KOH | 1924 | 0.95 / 0.87 | 0.70 | - | 4.7 | - | $H_2$ @ -196°C, 20 bar | Sangchoom et al.[129] (2015) |
| Starch + $CO_2$ | 3350 | 1.75 / 1.67 | 1.52 | - | 6.4 | 10.7 $mmol.g^{-1}$ | $H_2$ @ -196°C, 20 bar<br>$CH_4$ @ 25 °C, 20 bar | Li et al.[62] (2016) |

## 2.5 Templated carbons

Recently, many research attempts have been devoted to fabricating new carbon-based adsorbents with controlled and well-defined porosity inside the adsorbent's skeleton [130-132]. Traditional activation methods are associated with difficulties such as harsh reaction/operation conditions, disordered/ununiform/broad pore size distribution and relatively low carbon yield of final adsorbents. Templating approach is considered one of the effective strategies to produce porous carbons due to its ability for tailoring well-organized porous morphologies purposefully and precisely. This method makes it possible to adjust the pore structure and PSD of porous carbons precisely, facilitate the preparation conditions (i.e. reducing operating temperature and energy consumption), and improve the carbon yield, resulting from a low corrosion rate in the skeleton of carbon [131]. The advantages mentioned above make the templated carbons one of the most appropriate candidates for gas storage/separation in recent years [131-133].

Generally speaking, the templating approach involves several steps: (1) carbon raw materials and templating agents need to be mixed, where carbon is assembled with the templating agent (soft templating) or set inside the porosity of the agent (hard templating); (2) the resultant should be exposed to thermal treatment (i.e. carbonization) to remove the agent from the mixture [131]. Two main approaches of soft and hard templating are common to develop the porosity inside the carbon framework. Soft templating or endo-templating denotes the preparation of porous materials mainly using self-assembled organic molecules/supramolecules (e.g. surfactants in a particular solution and in the presence of a polymerizable carbon precursor [131]. They will be decomposed and removed from the media during the thermal condition. Consequently, the porous carbon materials can be directly obtained after the carbonization process (see Figure 9). The hard templating or exo-templating approach includes the following steps; (1) carbon precursor is introduced into the appropriate porous template agents (e.g. zeolites, and silica) through the methods of wet impregnation and chemical vapour deposition (CVD) or a combination of both of them, (2) the two phases appear to be polymerized and crosslinked and finally (3) suitable solutions such as HF or NaOH is used to decompose the inorganic template so that the well-ordered porous carbons will be obtained (see Figure 9) [134]. Between two templating techniques, hard templating is more economical and can provide higher specific surface area and ordered microporous architectures [13, 131].

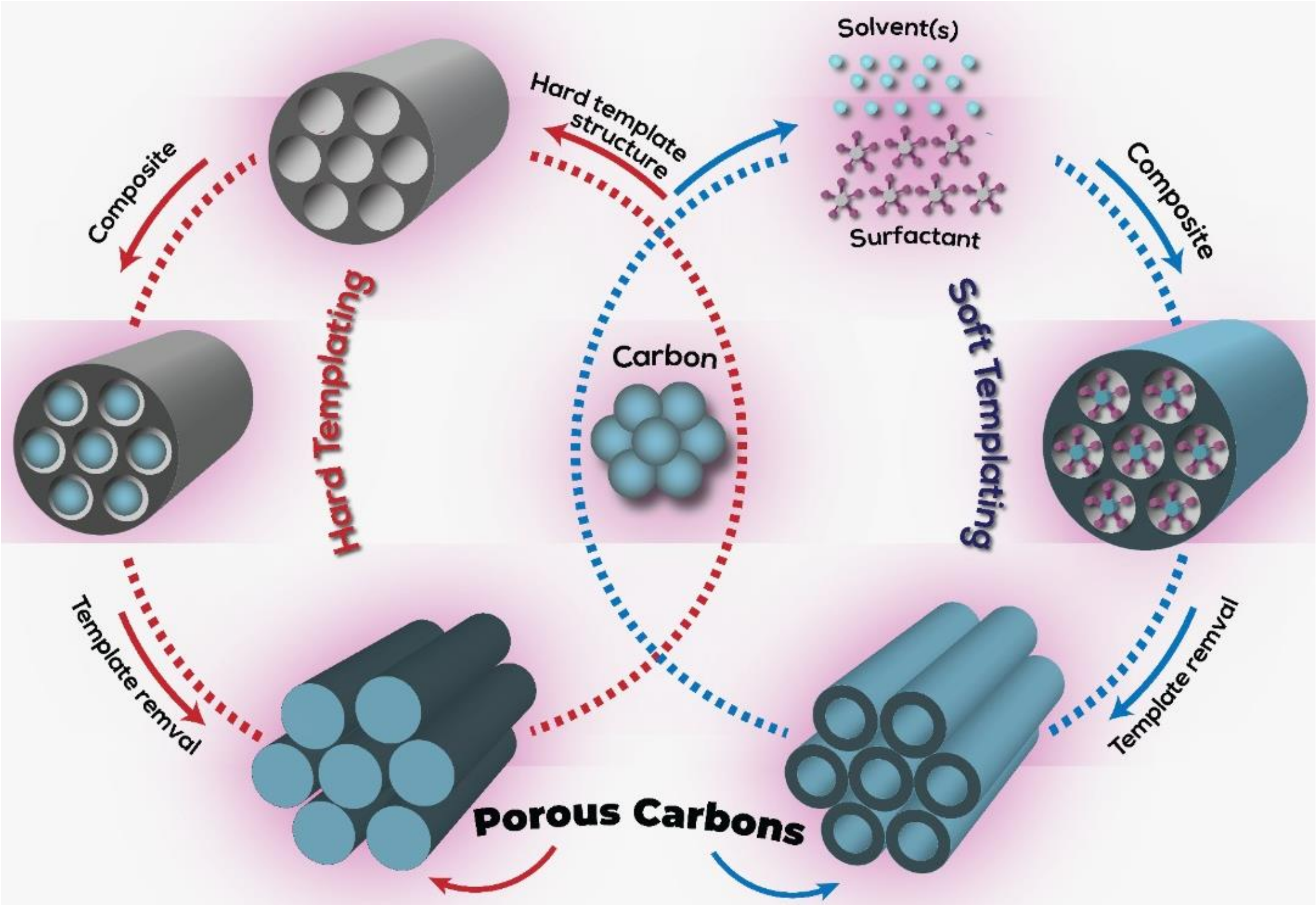

**Figure 9.** Schematic illustration of the synthesis process of soft and hard templating.

From a practical point of view, zeolite-templated carbons (ZTCs) are believed to be one of the critical subclasses of hard templating carbons, especially for gas storage uses. These microporous carbons represent high surface areas (up to 3500 $m^2.g^{-1}$) and a periodic array of pores complementary to the zeolite structure used in the template carbonization [133]. Due to the significant gas sorption capacity of ZTCs at room temperature, several researchers focused on the practical application of these adsorbents in the vehicle industry. For example, Stadie and co-workers successfully synthesized high template fidelity ZTCs to enhance $CH_4$ and hydrogen uptake at room temperature [133, 135]. The remarkably high gas storage capacity of ZTCs ($CH_4$: ~12 mmol.$g^{-1}$ at 25 °C and 35 bar & $H_2$: 8.27 mmol.$g^{-1}$ at 25 °C and 30 bar) was reported. An extensive surface area of 3591 $m^2.g^{-1}$ and, more importantly, the presence of narrow micropores (~1.2 nm) in the ZTCs structure was reported as the main reasons behind their significant storage capacity. Such a narrow pore width inside the materials' framework is nearly close to the optimal value of 1.14 nm, which is recommended for maximum gas uptake [7, 65].

The uniform crystal morphology of ZTCs is their other important characteristic, which provides the opportunity to have adsorbents with homogenius particle size distributions and higher packing density compared with the conventional activated carbons [136]. In recent research on $CH_4$ adsorption, Beta (BEA)[2] and faujasite (FAU)[3] zeolite structures were used as a solid template to prepare the ZTCs [136]. It has been demonstrated that the micropore size distribution of ZTCs was systematically controlled by post-synthesis thermal treatment. As a result, the unique feature of thermal contraction was observed for ZTC-types, which was not reported for former ACs. The final adsorbent with tailored micropore size distribution (in the range of 1.1-1.5 nm) indicated one of the most promising volumetric $CH_4$ storage values of 210 $cm^3.cm^{-3}$ within the pressure range of 0-65 bar. In a relevant study, an ordered carbon was synthesized via a soft-template approach as a potential adsorbent for $H_2$ adsorption [137]. The final adsorbent showed uniform pore size distribution with an average pore diameter of 6.26 nm and pore volume of 0.87 $cm^3.g^{-1}$. Hydrogen storage capacity was recorded in a volumetric apparatus at three different temperatures (-196, -

[2] BEA zeolite is among the most widely used zeolites. It exhibits a three-dimensional pore system formed by 12-membered ring channels with a diameter of 0.76 × 0.64 and 0.55 × 0.55 nm, which ensures good accessibility of acid sites, high thermal stability and high acidity.

[3] The faujasite framework has been attributed the code FAU by the International Zeolite Association. It consists of sodalite cages which are connected through hexagonal prisms.

78.5, and 25 °C), maximum gas uptake value of 1.27 wt.% was found to be reached at -196 °C and 1.05 bar pressure. Some of the ACs paprepared by templated method were listed in Table 6.

**Table 6.** Porous carbons prepared by template method for gas storage.

| Precursor + Template | Porosity | | | $H_2$ Uptake (wt.%) | $CH_4$ Storage Capacity ($V.V^{-1}$) | Measurement Condition | Reference |
|---|---|---|---|---|---|---|---|
| | $S_{BET}$ ($m^2.g^{-1}$) | Vt / $V_p$ ($cm^3.g^{-1}$) | $d_{ave}$ (nm) | | | | |
| Furfuryl alcohol + Zeolite $NH_4Y$ | 2135 | 1.4 / 0.57 | 1.50 | 2.24 | 10.98 $mmol.g^{-1}$ | $H_2$ @ -196°C, 1 bar<br>$CH_4$ @ 25 °C , 35 bar | Guana et al.[138] (2015) |
| Acetylene + Zeolite beta | 1073 | 0.75 / 0.29 | 0.90 | - | 80 | $CH_4$ @ 25 °C , 35 bar | Antoniou et al.[139] (2014) |
| Furfuryl alcohol + Zeolum zeolite molecular sieve | 3591 | - | 1.25 | 28.6 $mmol.g^{-1}$ | 12.32 $mmol.g^{-1}$ | $H_2$ @ -196 °C, 1 bar<br>$CH_4$ @ 25 °C, 35 bar | Stadie et al.[133, 135] (2012 & 2013) |
| Zeolite structure BEA + CVD | 2447 | 1.07 / 0.99 | 1.33 | - | 210 | $CH_4$ @ 25 °C , 65 bar | Choi et al.[136] (2018) |
| Zeolite structure FAU + CVD | 2145 | 0.95 / 0.88 | 1.33 | - | 185 | $CH_4$ @ 25 °C , 65 bar | Choi et al.[136] (2018) |
| Furfuryl alcohol + mesoporous $ZnO/Zn(OH)_2$ | 1013 | 3.39 / - | 0.67 | 1.14 | - | $H_2$ @ -196°C , 1.01 bar | Kim et al.[140] (2020) |
| Phloroglucinol + mesoporous $ZnO/Zn(OH)_2$ | 1075 | 3.01 / - | 0.74 | 1.19 | - | $H_2$ @ -196°C , 1.01 bar | Kim et al.[140] (2020) |
| Furfural alcohol + zeolite 13X | 3332 | 1.66 / 1.18 | 1.2 | 7.3 | - | $H_2$ @ -196 °C , 20 bar | Masika et al.[114] (2013) |
| Glucose + NaY zeolite | 1600 | - | 1.35 | - | 7 $mmol.g^{-1}$ | $CH_4$ @ 295 °C , 40 bar | Shi et al.[41] (2015) |

## 2.6 Carbide-derived carbons

Carbide-derived carbons (CDCs), or "mineral carbons", are attractive candidates for gas storage mainly due to their tunable pore size distribution (PSD) with sub-Angstrom accuracy [141, 142]. In some ways, many similarities exist between the synthesis of CDCs and the templating approach. The rigid metal carbide lattice acts as a template in the CDCs preparation, with the metal atoms extracted layer-by-layer. Thus, the as-prepared carbon structure is 'templated' by the carbide structure (see Figure 10). In this way, there is a significant chance of adjusting the porosity at the atomic level by controlling the chlorination condition (e.g. temperature) or selection of metal carbides (precursors) such as $B_4C$, ZrC, $Ti_3SiC_2$, $Ti_2AlC$ [143].

Structural properties of the initial carbides (precursors) and CDCs preparation method significantly impact the tunable PSD characteristics of these materials. CDCs can be produced by selective etching of metals from crystalline metal carbides [13]. Several approaches have been recommended for selective removal of metal or metalloid atoms from carbides, including hydrothermal leaching, thermal decomposition, etching at high temperatures in molten salts, halogenation and electrochemical etching at room temperature. Among them, halogenation is considered to be the most efficient route to prepare nanoporous carbons. Chlorine ($Cl_2$) is known as the most commonly used agent for halogens etching.

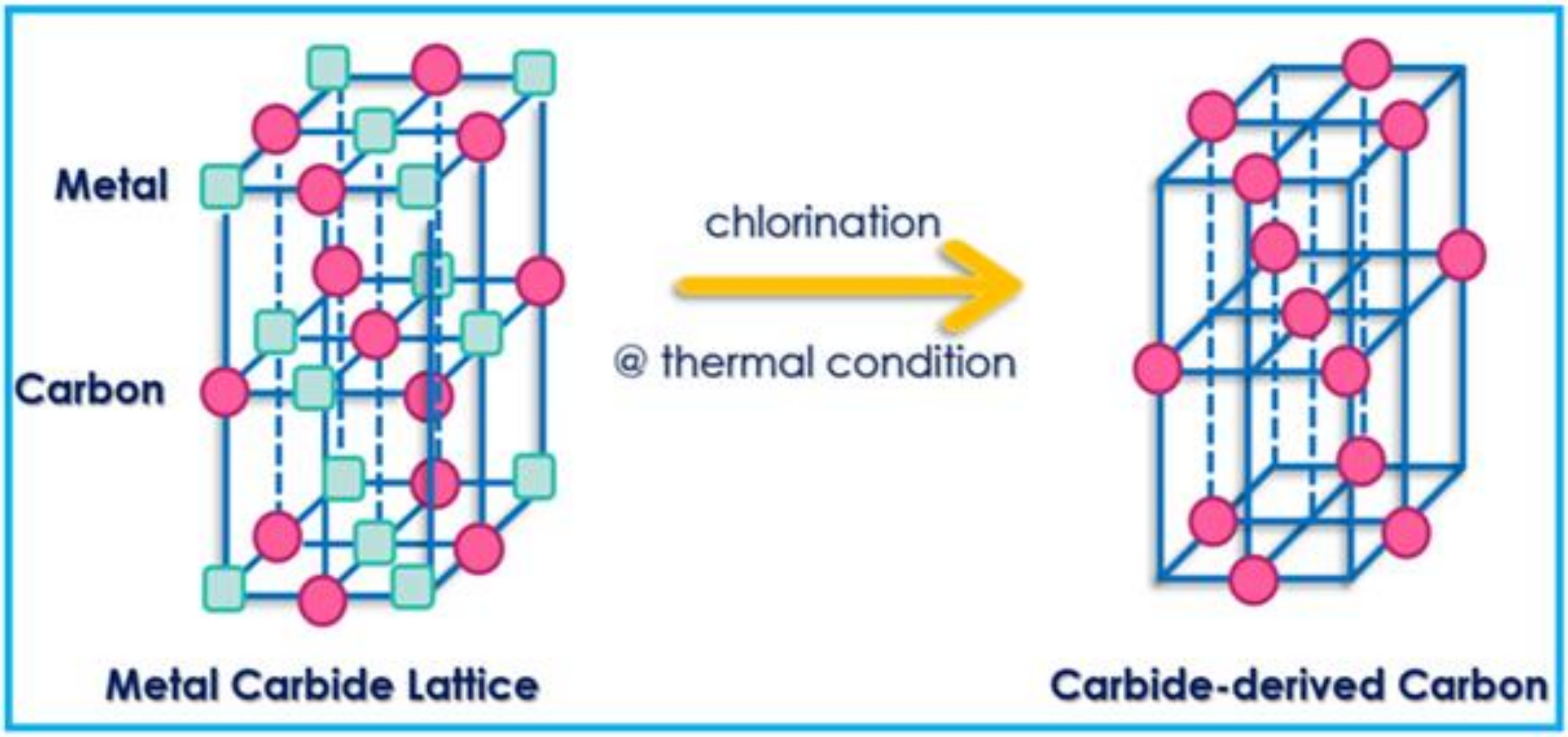

**Figure 10.** Schematic diagram of Carbide-derived carbons fabrication

As an example, titanium tin carbides ($Ti_2SnC$)-based CDCs were synthesized via chlorination at 400–1100 °C to study their gas adsorption properties [144]. The as-prepared CDCs under low chlorination temperature (400–500 °C) mainly showed an amorphous carbon structure contained with the chlorides. Increasing the temperature up to 600°C resulted in the disappearance of the chlorides; hence the main composition of CDCs was amorphous carbon. At higher chlorination temperatures (near 1100 °C), an apparent trend of graphitization was detected. The CDC chlorinated at 1100°C showed a $S_{BET}$ of 1580 $m^2.g^{-1}$ and a gas adsorption capacity of 206 $cm^3.g^{-1}$ (at 60 bar 25 °C) for $CH_4$ and 442 $cm^3.g^{-1}$ (at 35 bar -196 °C) for $H_2$.

As mentioned earlier, the pore size of CDCs can be fine-tuned by optimizing the preparation condition. However, there is a limitation in applying such materials, especially for gas storage or as electrochemical capacitors, where larger specific surface areas are required - the highest values of $S_{BET}$ reported for CDCs are in the range of 2000-2500 $m^2.g^{-1}$ [143]. Gogotsi and co-workers activated TiC based CDCs physically and chemically using air + carbon dioxide and KOH, respectively, under different thermal conditions. They investigated the influence of the activation on hydrogen and $CH_4$ storage capacity of the final products [145-147]. These early studies on post-synthesis activation of CDCs demonstrated the possibility of enhancing and adjusting the porosity of CDCs through the use of chemical or physical activation and the importance of the activation temperature and duration. After the activation, surface area and pore volume increased to 3360 $m^2.g^{-1}$and 1.4 $cm^3.g^{-1}$, respectively, two times the raw CDCs. The increase in total pore volume might be accompanied by the widening of small micropores and the formation of an additional mesopore volume contribution, which is not desirable for hydrogen storage. However, Gogotsi and co-workers found an enhancement in hydrogen uptake, reaching a value of 4.7 wt.% at -196 °C and 60 bar, up to 52% higher than the raw CDCs [147]. The point that should be remarked here is the benefit of higher porosity of the obtained carbons after activation for $CH_4$ uptake. Indeed, they observed 50 % enhancement in $CH_4$ uptake of activated CDCs (18 wt.%) compared to the raw samples (12 wt.%) [146]. Likewise, later work by Sevilla and Mokaya represented a comprehensive study on the possibility of improving the porous textural properties of CDCs via further post-synthesis treatments and the ramifications of such modifications on the hydrogen storage capacity of the adsorbents [143]. They found that post-synthesis chemical activation improves the CDCs' surface area and pore volume around 47% and 58%, respectively. Consequently, the hydrogen storage capacity was promoted up to 63%.

## 2.7 Ordered (structured) adsorbent

Traditional carbon-based adsorbents are fabricated in beads or granules, which are associated with some disadvantages in operational conditions such as mass transfer limitations and loss of the space between the particles, affecting the bed's volumetric capacity [148]. It is possible to tackle the problem by reducing the particle size to less than 0.5 mm or using the powder form of the adsorbent. However, applying smaller particle size adsorbents causes a higher pressure drop and bed's temperature (due to higher heat of adsorption) [149]. With advantages such as a low bed's pressure drop, good thermal properties, fair mass transfer kinetics and sufficient working capacity, structured adsorbents were reported as promising alternatives for conventional adsorbents in the form of beads or granules [148, 149]. Ordered/structured porous adsorbents can mainly be categorised into dense structures (e.g. monoliths and pellets), foams, and fibres, which will be investigated in this section.

### 2.7.1 Dense carbons

Mobile adsorptive gas storage/separation requests need to consider the volumetric adsorption capacity of microporous carbon materials besides their gravimetric capacity. The gravimetric storage capacity of adsorbents is strongly affected by their microporosity; however, the volumetric capacity depends on both porosity and packing density of the as-made adsorbents. Although ACs in powder form show a large specific surface area (>3500 $m^2.g^{-1}$), their low bulk density might question their ability for energy storage. Reshaping the low-dense ACs powder into dense/bulked structures such as pellets or monoliths makes it possible to compensate for the limiting factor [150]. The rational design of monoliths to carbon-based adsorbents can simultaneously achieve optimal bulk density and microporosity, which is genuinely convenient for practical applications.

Activated carbons can be compacted into monoliths or pellets with or without the use of a binder simply by the selection of appropriate carbon precursors. The synthesis procedures to prepare bulked ACs with/without binder are represented in Figure 11 [21]. It needs to be remarked that besides the templated and carbide-derived carbons, dense ACs are also promising candidates for gas storage. With relatively high volumetric storage capacity, these materials enjoy simple, well-established and low-cost production methods [112].

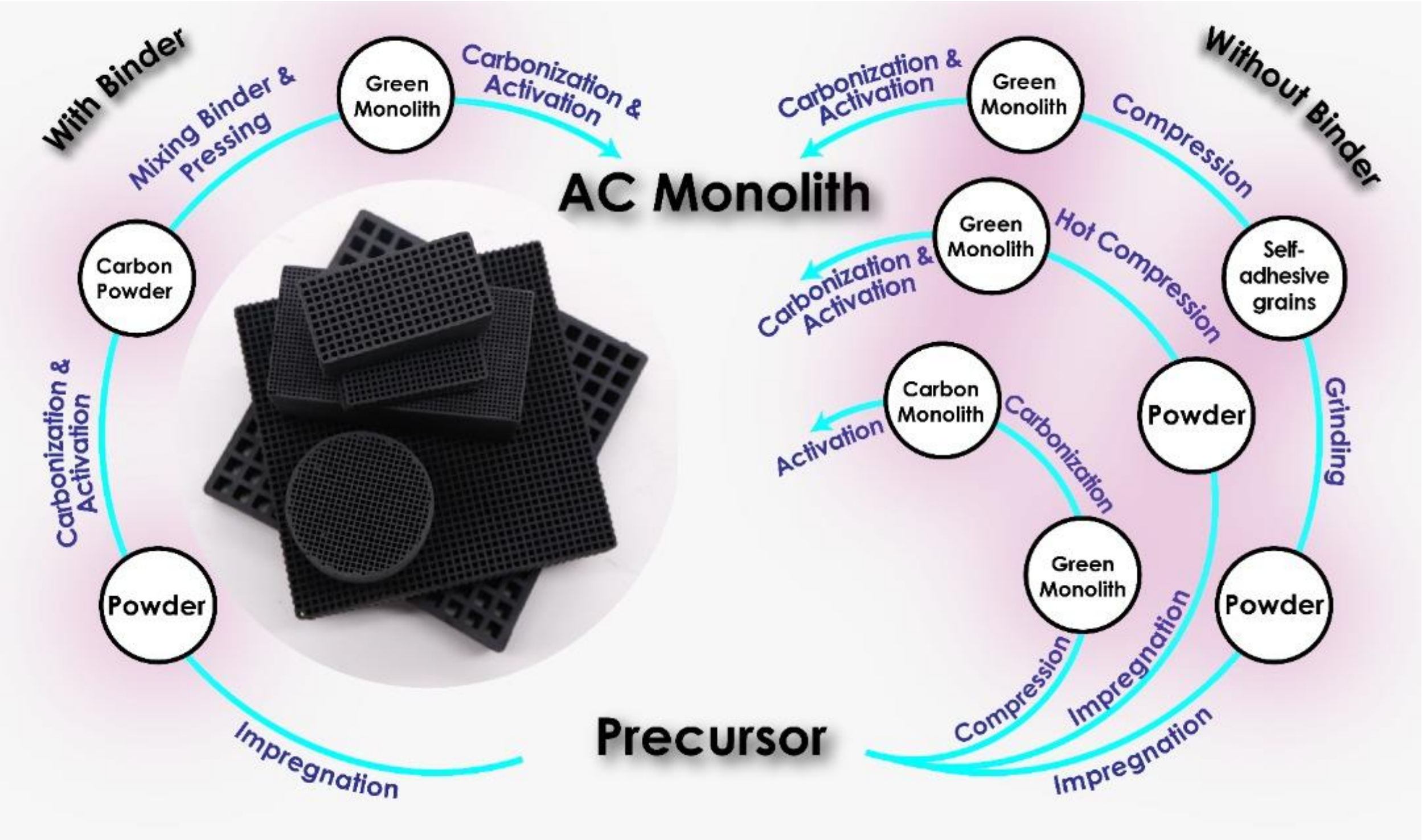

**Figure 11.** Synthesis procedures of dense adsorbent with/without binder

One effective way to prepare dense ACs is the compactification approach that includes a pressing step before thermochemical activation [97]. Adeniran and Mokaya represented a method wherein powdered KOH and carbon precursor mixtures were mechanically compressed under 7400 bar pressure into pellets/disks. Afterward, the compressed mixture was activated at two different temperatures, 600 °C (low levels of activation) and 800 °C (high levels of activation). As-made bulked ACs possessed higher porosity than analogous conventionally activated carbons. The designed method showed good efficiency under either low or high activation temperatures. Highly compressed-activated carbons treated at 800 °C showed a surface area of around 4000 $m^2.g^{-1}$ and pore volume up to 3.0 $cm^3.g^{-1}$ along with superior hydrogen uptake (9.6 wt.% or 38 $g.l^{-1}$ at -196 °C and 40 bar). In a similar study, the effect of densification was studied on the porosity and gas storage properties of the corresponding ACs [97]. The results indicated that the produced carbons held a significant proportion of their porosity (3500 $m^2.g^{-1}$ and 2.7 $cm^3.g^{-1}$) whilst attaining exceptional volumetric hydrogen storage (49 $g.l^{-1}$ at -196 °C and 40 bar).

Chen and colleagues recently reported the development of high-performance dense carbon nanoflower pellets (Figure 12), via a facile method of free radical polymerization of acrylonitrile, for on-board $CH_4$ storage applications [150]. Old-style efforts have been conducted by hydraulic pressing the AC powders into dense pellets with an almost 10% binder. In this way, fragile pellets with low mechanical stability were hardly made. In a particular way designed by Chen et al., freestanding polymer pellets can be readily made via controlling the packing of AC nanoflowers monodispersed particles, which significantly improves the bulk density. Freestanding polymer pellets can be readily obtained by reducing the solvent concentration (from 2:1 to 1:2 acetone/acrylonitrile) before the polymerization step and drying the formed polyacrylonitrile particles at 70 °C under vacuum. As-made adsorbents showed a record-setting bulk density (0.96 $g.cm^{-3}$) with interconnected micro/meso/macropores and a surface area of 1077 $m^2.g^{-1}$. This simple production method could record high $CH_4$ storage capacities of 165 and 196 $cm^3.cm^{-3}$ at 35 and 65 bar, respectively.

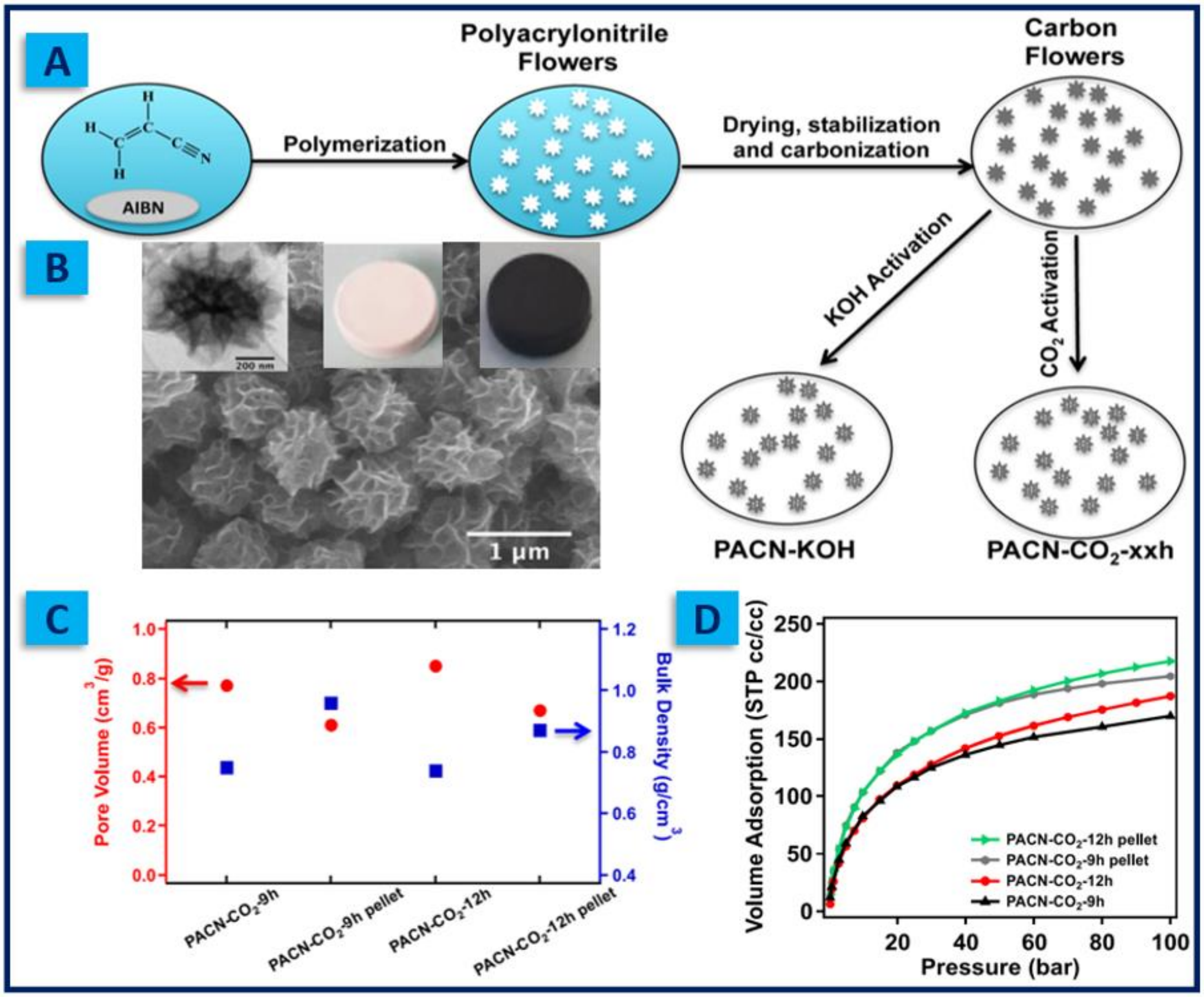

**Figure 12.** (A) Synthetic scheme of porous carbon nanoflowers powders with different activation methods, (B) SEM image of the samples, (C) Summary of porosity and bulk density for activated carbon nanoflower pellets vs powders, (D) Volumetric $CH_4$ uptakes at 25 °C for activated carbon nanoflower pellets vs powders. Reproduced from ref [150]. Copyright 2020, American Chemical Society.

### 2.7.2 Foam-like carbons

Carbon foams are spongy carbonaceous materials with cellular microstructures and hierarchical pore arrangements that attracted significant interest because of their unique structure and extraordinary performance in adsorption, energy storage, catalyst support, sensors and electromagnetic interference shielding applications [151]. Carbon foams possess adjustable 3D porous architecture with high strength, good thermal stability, high electrical conductivity, and large specific surface area. As illustrated in Figure 13, carbon foams hold two kinds of pore structures; (i) open-cell structures and (ii) closed-cell cellular structures. The first one includes ligaments and joints inside a unit cell (Figure 13 (A)), while the other type is composed of the cell walls, ligaments and joints. The ligament denotes the edge of the cell which connects the two neighbouring cells; more information about carbon foam structures was reported elsewhere [151].

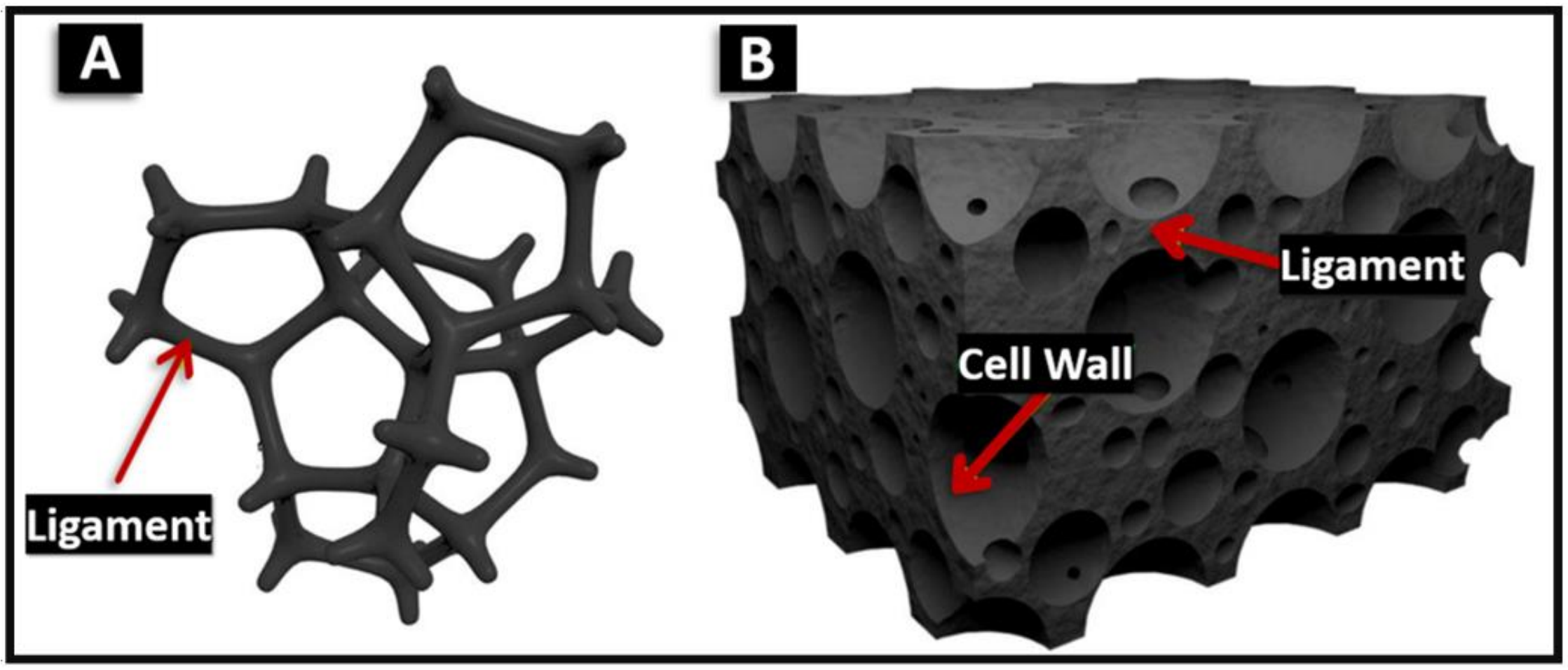

**Figure 13**. Modelling of the typical microstructure of carbon foam: (A) open-cell, (B) closed-cell. Reproduced from ref [151]. Copyright 2020, Royal Society of Chemistry.

Carbon foams can be produced using different carbonaceous precursors such as polymers [152, 153], pitches [46, 154], and biomass materials [155, 156]. Generally speaking, for carbon foam fabrication, regardless of the precursors, carbonization step at 750-1400 °C is needed to remove impurities and un-wanted atoms (i.e. H, S, N, and O) and provide the final product with a better structure and morphology of carbon matrix [151].

A comprehensive study on the adsorption of a series of gasses, $N_2$, $CH_4$, and $CO_2$, on carbon foams, was reported by Arami-Niya et al. [156]. They produced nitrogen-doped AC foams from banana peels using a self-template method with zinc nitrate, 2-aminophenol-and furfural and investigated the effect of carbonization temperature and post-carbonization, $CO_2$ activation on the pore structure and gas uptake of the final adsorbent. The carbon foams demonstrated a nitrogen content of about 6.0 wt.% and featured cellular macroporous structures with BET-specific surface areas up to 1426 $m^2.g^{-1}$. The potential of the carbon foams for $CO_2/N_2$, $CO_2/CH_4$ and $CH_4/N_2$ separations was assessed using adsorption isotherm at 40 bar and 298 K. The results showed gas uptake values of 3.29, 5.29 and 9.21 $mmol.g^{-1}$ for $N_2$, $CH_4$ and $CO_2$ respectively.

Pitches are considered a customary carbonaceous precursor used to produce carbon foams due to their low cost and ease of accessibility [46, 157]. Typically, the synthesis process of pitch-based materials involves several steps of foaming, carbonization, graphitization and finally activation. Foaming strategies include blowing, and spontaneous foaming, wherein pitch releases light components and small molecules (volatilization). Afterwards, as-made foam materials experience carbonization (750-1400 °C) and graphitization (usually higher than 2100 °C), to form the final carbon foam structures [151]. In similar research by Arami-Niya and coworkers, AC monoliths were fabricated with foam-like features from petroleum tar pitch at ambient pressure using coal powder to stabilize the liquified pitch [46]. They studied the impact of coal to pitch ratio on foam morphology and micropore development. The carbon-foam sample with 50 wt.% coal to pitch featured an open-cell structure with cell widths of around 2 mm and a well-developed microporosity that presented a BET specific surface area of 1044 $m^2.g^{-1}$. Moreover, this sample showed adsorption capacities of 7.398 $mmol.g^{-1}$ $CO_2$, 5.049 $mmol.g^{-1}$ $CH_4$ and 3.516 $mmol.g^{-1}$ $N_2$ at 25 °C and pressures close to 35 bar.

### 2.7.3 Carbon fibres

Over the past few decades, activated carbon fibres (ACFs) attracted lots of interest due to their unique characteristics comprising fibrous structure[158], well-developed porosity[159, 160], high volumetric gas adsorption capacity[160], excellent packing density[161], fast adsorption kinetics and simplicity of handling[158, 160, 162]. These materials can be synthesized using different raw materials (like phenolic resins, mesophase pitch, pitch fiber, polyacrylonitrile, or biomass). However, due to the extra processing steps, e.g. converting the initial materials into the fibrous form, production of these materials is relatively expensive [160]. The main challenge behind the fabrication of ACFs is to obtain homogeneous pore size structures which can improve their gas adsorption properties.

ACFs can be produced through carbon fibres' physical or chemical activation [158]. No matter of the precursors, e.g. polymers, pitches or biomass, the production method of ACFs includes the steps of spinning, stabilization, pre-treatment (acid impregnation before carbonization), carbonization and activation [158, 160]. During the spinning process, the first step of the ACF production, the carbon-based precursors in the shape of granular or powder, will be converted into a continuous fibre. Diverse techniques have been reported for the spinning process, such as melt spinning [158, 163, 164], wet spinning [165] and dry spinning [166]; depending on the nature of the precursor, each of these methods can be chosen. Melt spinning involves three steps of (1) melting the precursor, (2) extruding through a spinneret containing numerous small capillaries, and finally (3) drawing of the fibres. During the wet spinning, fibres are produced from the precipitation of a concentrated solution, extruded inside a coagulation fluid. In the dry spinning method, the fibres are formed by evaporating the solvent in a drying vessel.

Oxidative or thermal stabilization is considered a crucial phase to reach high-quality ACFs. The treatment temperature in the range of 150-400 °C, precursor diameter, and raw fibre characteristics are the critical thermal treatment parameters [167]. The primary purpose of stabilization is to produce a fibre structure that can withstand the rigours of high-temperature processing. Carbonization comprises several complex physiochemical changes such as dehydrogenative condensation, polymerization, aromatization and removal of almost every impurities (e.g. O, N, CO, $H_2$ and $H_2O$) from the as-prepared fibre resulting in up to 90% improvement in carbon content [167]. Activation was detailed described earlier in section 2.

Conte and co-workers synthesized a series of ACFs via $CO_2$ activation of poly p-phenylene terephthalamide, commercially known as Kevlar[4] [160]. They assessed the porosity and the gas adsorption properties of the as-prepared samples in various activation conditions, including time (in the range of 60-240 min), temperature (in the range of 750-850 °C) and $CO_2$-flow rate (0.3, 0.9 and 1.2 $l.min^{-1}$). Under the optimum activation condition, the ACFs illustrated a high micropore fraction up to 94% of the total pore volume with a specific surface area of 1109 $m^2.g^{-1}$ and a $CH_4$ uptake of 5 $mmol.g^{-1}$ at 40 bar and 25 °C. In a similar study, Hwang et al. developed a new strategy to control the surface porosity of activated carbon fibres and maximise hydrogen storage capacity [168]. They prepared KOH-activated carbon fibres from polyvinyl alcohol/polyacrylonitrile (PVA/PAN) bi-component, as a precursor, using the wet spinning method. As- prepared ACFs showed a highly porous structure with a surface area of 3058 $m^2.g^{-1}$ and micropore volume of 1.18 $cm^3.g^{-1}$. AFCs were measured to have a superior hydrogen storage capacity of 5.14 wt.% at -196 °C and 100 bar, much higher than the previously reported ACFs. In another study, Kostoglou et al. developed a mechanically stable and flexible nanoporous carbon from almost new attractive starting materials of raw viscose rayon fabric [92]. The rayon fabric was first impregnated in an aqueous solution of 4.0 wt.% ammonium chloride and 4.0 wt.% zinc chloride. After drying, the sample was introduced into a tubular furnace to be activated by heating at 930 °C under the $CO_2$ atmosphere (see Figure 14). The values of $S_{BET}$ and $V_{micro}$ of the prepared ACs were estimated to be ~1205 $m^2.g^{-1}$ and ~0.47 $cm^3.g^{-1}$, respectively. The prepared adsorbents were evaluated for three important applications, including (i) gas storage, (ii) selective separation of $CO_2$ over $CH_4$ based on the IAST model and (iii) electrochemical energy storage in supercapacitor devices. In the $H_2$ and $CH_4$ uptake case, the fabric-based carbon showed acceptable values of 1.68 wt.% (at -196 °C & 1 bar) and 0.72 $mmol.g^{-1}$ (at 25 °C and 1 bar), respectively.

---

[4] Poly p-phenylene terephthalamide (PPTA) namely Kevlar® is an aramidic synthetic fiber with a high content of carbon and, typically, high mechanical properties as resistance to traction, cut, heat and flames. The chemical formula of Kevlar® monomer is $[-CO-C_6H_4-CO-NH-C_6H_4-NH-]_n$

[160] G. Conte, S. Stelitano, A. Policicchio, F.D. Minuto, V. Lazzaroli, F. Galiano, R.G. Agostino, Assessment of activated carbon fibers from commercial Kevlar® as nanostructured material for gas storage: Effect of activation procedure and adsorption of CO2 and CH4, Journal of Analytical and Applied Pyrolysis 152 (2020) 104974.

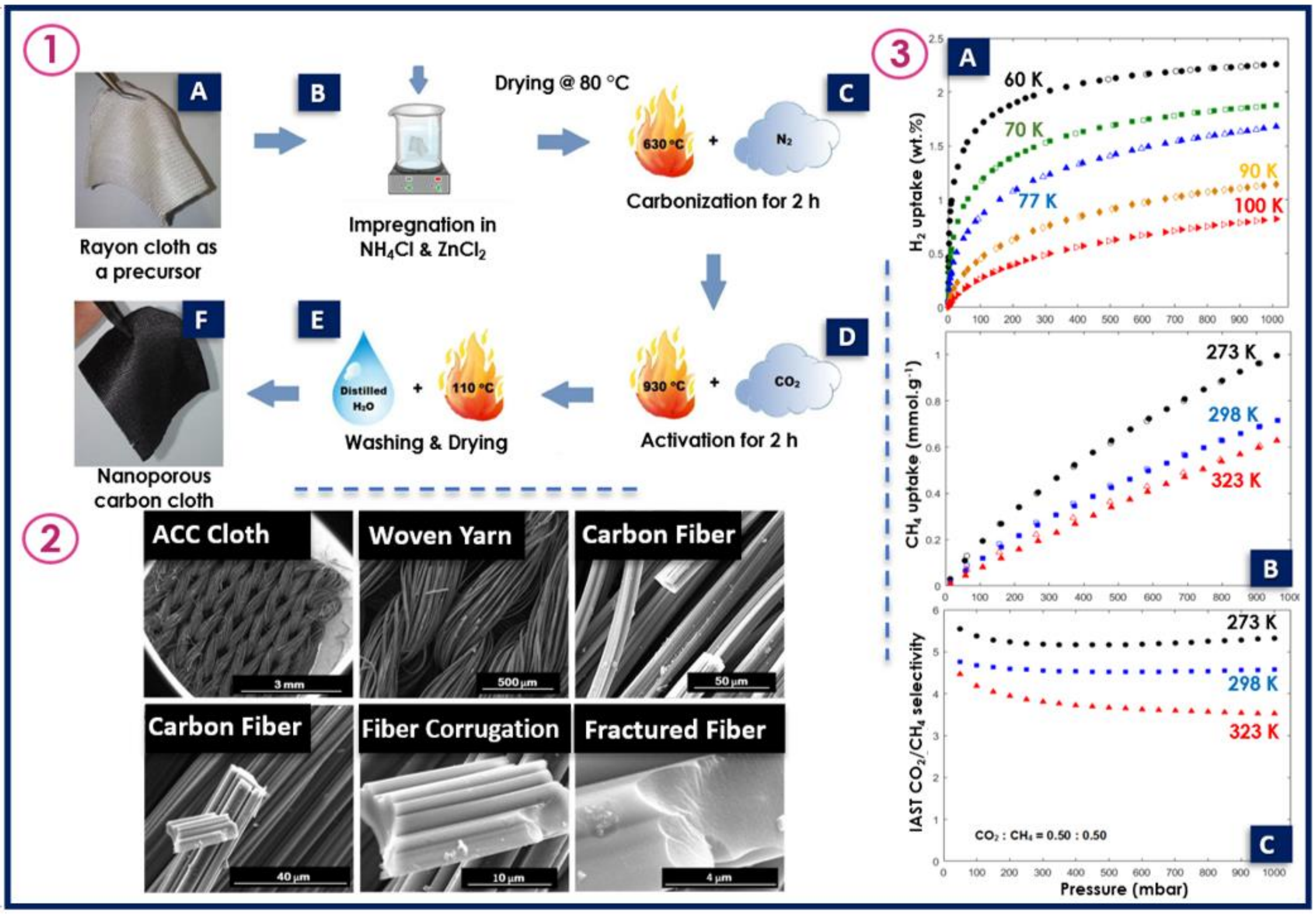

**Figure 14**. 1) Synthesis of the nanoporous material, carbonization and $CO_2$ activation procedures. 2) SEM images at different magnifications showing ACC cloth, woven yarns, carbon fibers, fiber corrugations and fractured fiber surface. 3) $H_2$ and $CH_4$ adsorption (solid symbols) and desorption (open symbols) isotherms together with $CO_2$ over $CH_4$ gas selectivity plots. Reproduced from ref [92]. Copyright 2017, Elsevier Ltd.

## 3. Carbon Nano-allotropes

ACs possess amorphous nature, disorderliness inside their framework and complicated crystalline structures originating from defects, curvatures, edges, heteroatoms or even randomly arranged stacked-carbon fragments [169]. However, carbon nano-allotopes family is a highly ordered material organized in particular carbon unit cells interconnected to adjacent cells [170]. The exceptional capability of carbon atoms to participate in robust covalent bonds with other carbon atoms in various hybridization states (sp, $sp^2$, $sp^3$) enables them to create a wide range of nanostructures, from small molecules to long chains [169]. As a result, as demonstrated in Figure 15, different types of ordered nanostructure carbons of 0D carbon fullerene, 1D carbon nanotube, 2D graphene and 3D graphite, can be formed. Desirable properties such as high specific surface area, superior electron mobility, easy self-assembly with controlled microstructures, high mechanical, chemical, thermal, and electrochemical stability make these ordered adsorbent promising candidates for gas sorption, separation and storage. The following section discussed these novel structures and their gas adsorption properties.

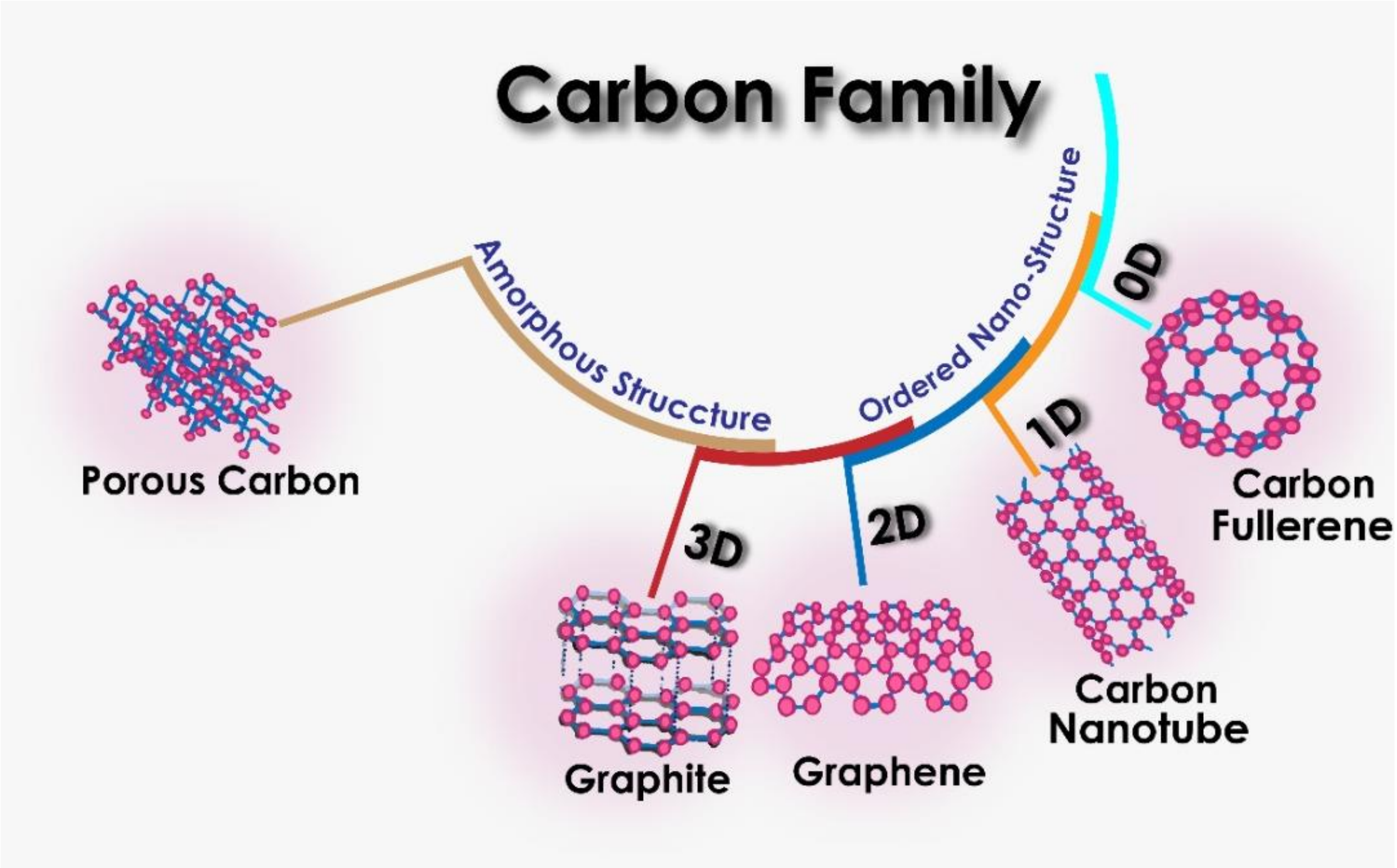

**Figure 15.** Different classes of porous carbons family

### 3.1 0D Carbon fullerenes

The discovery of carbon nano-allotrope structures began with the detection of fullerenes (e.g. $C_{60}$), known as the smallest carbon nanostructure and a representative of 0D carbon nano-allotrope [171]. Besides $C_{60}$ molecule, several other fullerenes such as $C_{20}$, $C_{70}$, $C_{82}$ and $C_{84}$ were subsequently discovered and introduced by researchers [169, 172]. However, most relevant studies focus on the synthesis and applications of $C_{60}$ [173-176].

Fullerenes have closed-cage structures composed of $sp^2$ hybridized carbon atoms connected by single and double bonds. Each $C_{60}$ molecule comprises 60 $sp^2$ carbon atoms organized in a series of hexagons to develop diverse structures such as a hollow sphere, ellipsoid, tube, or any other configuration. The most famous member of this family is buckminsterfullerene ($C_{60}$)[5]. The closed fullerenes, especially $C_{60}$, which is so-called buckyballs, have the shape of a soccer ball with twelve pentagons and twenty hexagons in its particular framework (Figure 16).

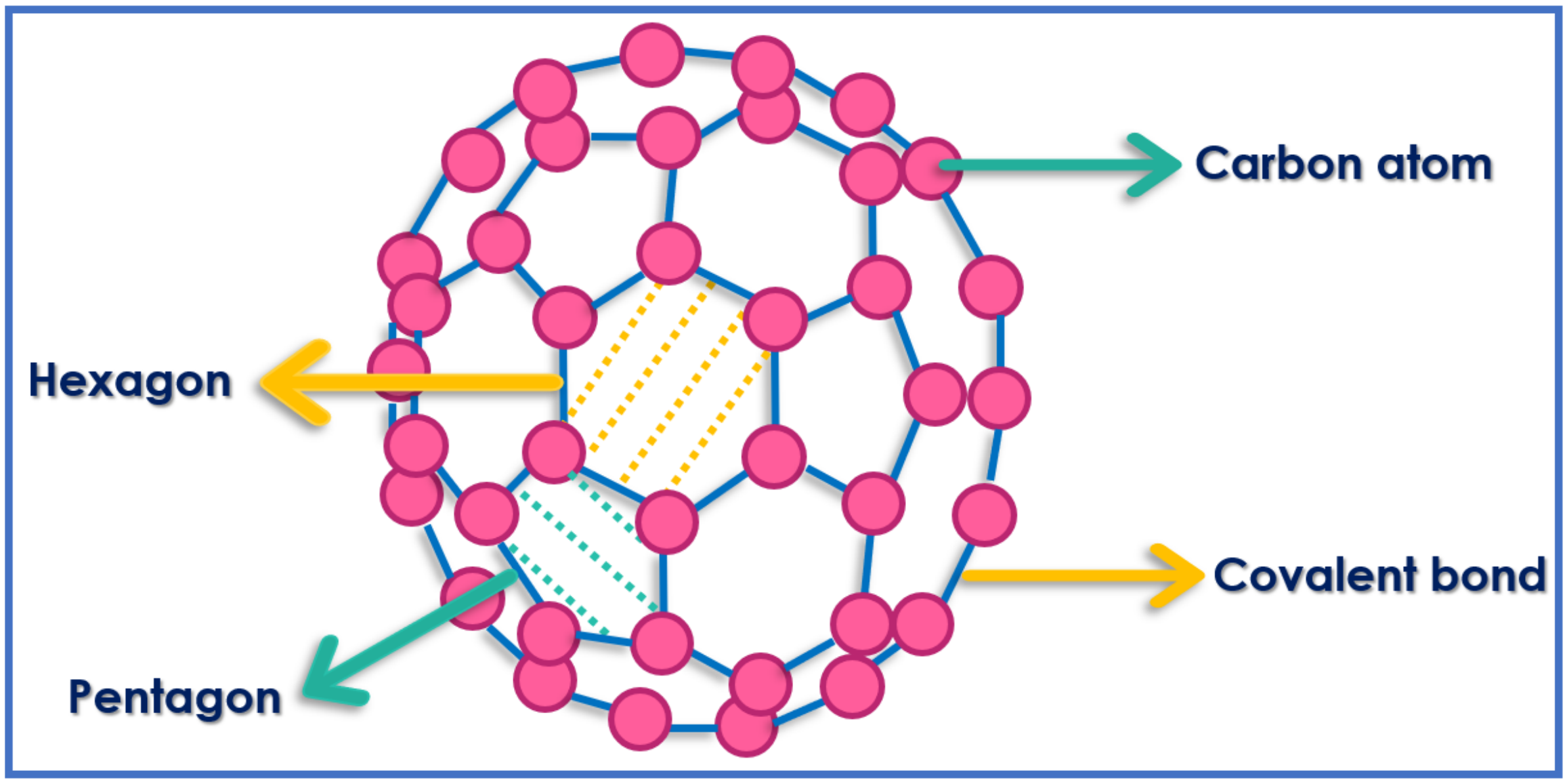

**Figure 16.** Configuration of buckyballs $C_{60}$

Generally, carbon fullerenes (i.e. $C_{60}$, $C_{70}$ and other configurations) are synthesized via two basic synthesis methods, including (i) vaporization through the arc and plasma discharges [171, 172, 177-179] and (ii) laser irradiation [171, 180]. The arc method, also called plasma arcing, is the first and most convenient way to fabricate fullerenes in reasonable quantities; the process is conducted using an electric current across two carbon-based electrodes under inert gas conditions.

---

[5] The buckminsterfullerene ($C_{60}$) was discovered in 1960 by Kroto et al. and named in honor of the famous architect Buckminster Fuller.

Fullerenes are created in the appearance of soot by plasma arcing of carbonaceous materials, especially graphite, as raw material. One drawback associated with the technique is producing a complex mixture of components, so additional purification is recommended to separate fullerene from the mixture [179, 181]. The laser irradiation method involves several steps; first, the sample becomes ready using laser evaporation of graphite rods with a catalyst mixture (ratio of 50:50) at 1200 °C under argon atmosphere, then thermal treatment in a vacuum condition is applied, and finally, purification will be conducted to remove the fullerene from carbon deposits (i.e. soot)[181].

Moreover, alternative preparation techniques such as naphthalene pyrolysis[182] and hydrocarbon combustion [183, 184] have been proposed; the last one is useful for the extensive commercial production of fullerenes. The associated drawbacks of most fullerene preparation techniques are (1) low yields and (2) difficulty of isolation/purification of the final products. As a result, although carbon fullerene originates from cheap and abundant raw materials such as graphite, its production is a high-cost process with undesired environmental impacts.

Fullerene's curved geometry helps these structures to store guest species either outside their spherical surface, known as exohedral adsorption, or inside the nano-cages cavity called endohedral adsorption or encapsulation. $C_{60}$ molecules can form monolayer and multilayer films of gas adsorbate inside their particular framework. Methane and hydrogen adsorption on the $C_{60}$ film seems to yield well due to the large curved C60 molecules (1 nm diameter), with a highly corrugated adsorption potential [185].

So far, gas adsorption properties of the fullerene-based adsorbents have been mostly explored through theoretical studies [175, 186-190], and limited experimental information is reported in the literature [191, 192]. Saha and Deng proposed an effective process to truncate $C_{60}$ buckyball molecules partially and create suitable pore textures for hydrogen adsorption by controlled oxidation at 400 °C and 2 bar oxygen pressure [191]. They made “real pores” within the carbon fullerene by rupturing the $C_{60}$ structure, and opening closed cages. The hydrogen storage capacity of pure and ruptured $C_{60}$ samples was measured using a magnetic suspension microbalance at different temperatures of -196, -130 and -50 °C and pressure ranges of 120 and 150 bar. The results indicated that the gas adsorption capacity of truncated fullerene molecules at -196 °C at 120 bar was enhanced more than three times from 3.9 wt.% to 13 wt.% compared to the pristine $C_{60}$ structure. Such an adsorption capacity strongly suggests that partial oxidation has opened the fullerene cages and made the buckyballs' inner surface area accessible to hydrogen molecules. The increased density of the partially truncated fullerene sample(2.52 $g.cm^{-3}$) compared to that of pristine fullerene (1.97 $g.cm^{-3}$) measured by helium pycnometery confirmed accessibility of gas molecules to the internal volume of the partially truncated C60 fullerene after opening the cages.

The robust spherical cage-like structure of fullerenes theoretically provides ample space for capturing/storing certain gases such as $H_2$ and $CH_4$. However, the challenging question for onboard gas storage application of these materials is the gas accessibility and the presence of active and selective adsorption sites in their spherical structure. Zottel et al. investigated available adsorption sites for $CH_4$ storage on fullerenes $C_{60}$ by a combined theoretical and experimental analysis [192]. They measured the adsorption capacity of different sites by mass spectrometry analysis, density functional theory and molecular dynamics (MD) simulations. They distinguished four different adsorption sites, including (1) registered sites above the carbon hexagons and pentagons, (2) groove sites between adjacent fullerenes, (3) dimple sites between three adjacent fullerenes, and (4) exterior sites.

Besides, it should be noted that confinement of a guest in the cavity of a fullerene cage would affect both the host and the guest molecules. Lately, the stability of "endohedral fullerenes" (so-called gas@$C_{60}$), where guest gas molecules are enclosed inside the fullerene cage, has been investigated experimentally and theoretically [173-175, 193-195]. The term "endohedral" originates from a combination of Greek words "endon" – within, and "hedra" – face of a geometrical figure [195]. Endohedral fullerenes can be formed in different ways; however, one effective way might be forming the fullerene in the presence of the endohedral species under high temperature and pressure [195]. Mohajeri et al. studied the $CH_4$ storage capacity of three different carbon fullerenes ($C_{60}$, $C_{84}$ and $C_{180}$) as well as the stability of these structures with confined $CH_4$ molecule(s)[175]. They analysed the empty cages and the formed complexes stability via electronic properties of the carbon fullerenes before and after $CH_4$ encapsulation based on first-principle density functional theory (DFT) calculations (see Figure 17). The binding energy values proved that $C_{60}$ (0.638 eV) and $C_{84}$ (0.559 eV) cages with one confined $CH_4$ molecule were stable. In the case of $C_{180}$ fullerene, the most stable configuration[6] was found when six $CH_4$ molecules were involved with the complex. To investigate the mechanical strength of $C_{60}$ and $C_{84}$ and their complexes, Mohajeri et al. performed Young's modulus analysis, which characterises how easily the structure can stretch and deform [175]. The results indicated that the strength of fullerenes with confined $CH_4$ molecules drastically decreases. In a similar study, the stability of endohedral n$CH_4$@$C_{60}$ was investigated through the combined DFT-based computation methods [173] and it was shown that the deposition of one $CH_4$ molecule in the spherical fullerene might be the only stable complex that ever formed for the $C_{60}$ cages.

[6] The most stable configuration of gas@fullerene complex demonstrate the geometry with the lowest mean polarizability.

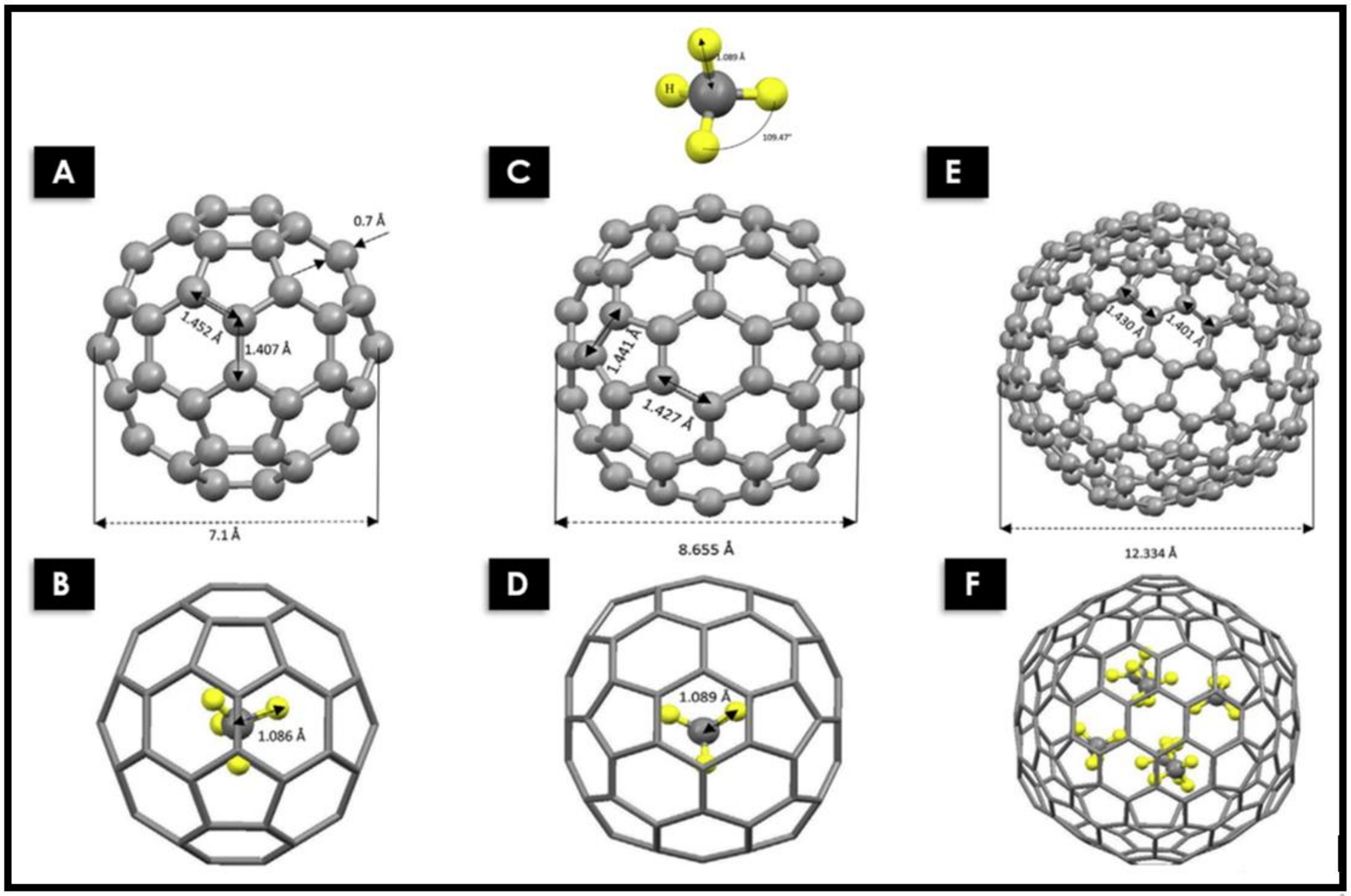

**Figure 17.** Optimized structures of $CH_4$ as well as A) $C_{60}$, B)$1CH_4@C_{60}$, C) $C_{84}$, D) $1CH_4@C_{84}$, E) $C_{180}$ and F) $6CH_4@C_{180}$ calculated by DFTB + code. Reproduced from ref [175]. Copyright 2018, Elsevier Ltd.

One of the other approaches for forming endohedral fullerenes is “molecular surgery”, wherein an orifice will be created on the nanocage to make open-cage fullerene accessible to atoms or small molecules (Figures 18 and 19). After a series of chemical reactions, the orifice becomes smaller and eventually closed to isolate the guest species [193, 194]. The orifice's opening size might be considered a limiting factor that restricts the insertion of the guest species inside the nanocage. As shown in Figure 18, in the synthesis of $H_2@C_{60}$, Komatsu et al. proposed a four-step organic reaction (A-D) in which a 13-membered ring orifice of an open-cage fullerene gently reduced and finally closed [194]. During each of these steps, complete retention of encapsulated $H_2$ was confirmed by observing the characteristically upfield-shifted NMR signal of the incorporated $H_2$. The stability of the endohedral fullerene $H_2@C_{60}$ was assessed, and the results proved that the complex was nearly as stable as $C_{60}$ itself; for instance, the encapsulated $H_2$ did not escape even when the temperature rose up to 500 °C for 10 min.

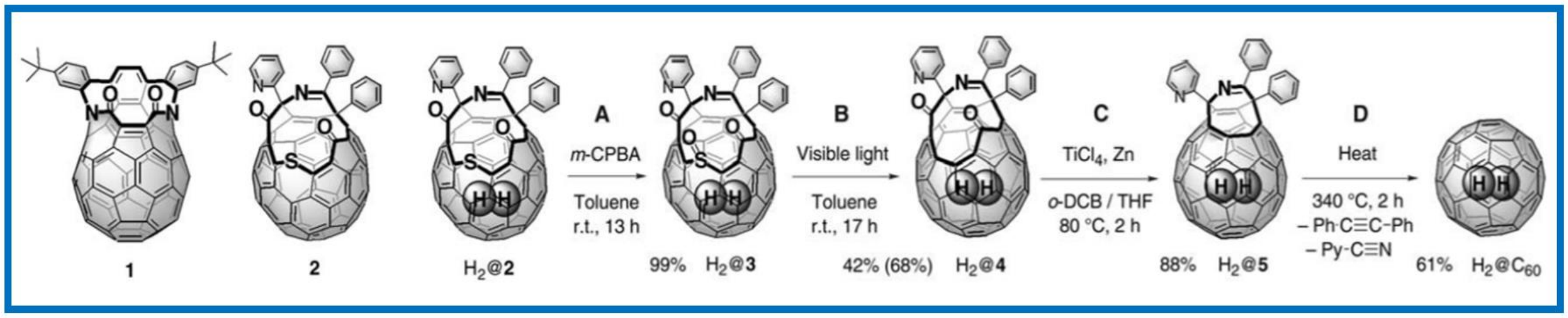

**Figure 18.** Size reduction and closure of the orifice of the open cage fullerene encapsulating hydrogen in a four-step process. Percentage values are product yields; shown in parenthesis is based on the consumed precursor. m-CPBA, r.t., and o-DCB stand for m-chloroperbenzoic acid, room temperature, and o-dichlorobenzene, respectively. Reproduced from ref [194]. Copyright 2005, American Association for the Advancement of Science.

In a similar study, Bloodworth and co-workers fabricated an endohedral fullerene of $CH_4@C_{60}$ [174]. They successfully encapsulated a single $CH_4$ molecule, as a large organic molecule, inside the $C_{60}$ fullerene cage (Figure 19). Also, the vital stage of the orifice contraction was conducted via a photochemical desulfinylation of an open fullerene, even though the presence of the endohedral molecule significantly inhibits the process. The $CH_4@C_{60}$ was characterized by high-resolution mass spectrometry, NMR spectroscopy and X-ray crystallography. The $^1H$ and $^{13}C$ spin-lattice relaxation times for endohedral $CH_4$ were measured and it was shown that $CH_4$ is freely rotating inside the $C_{60}$ cage.

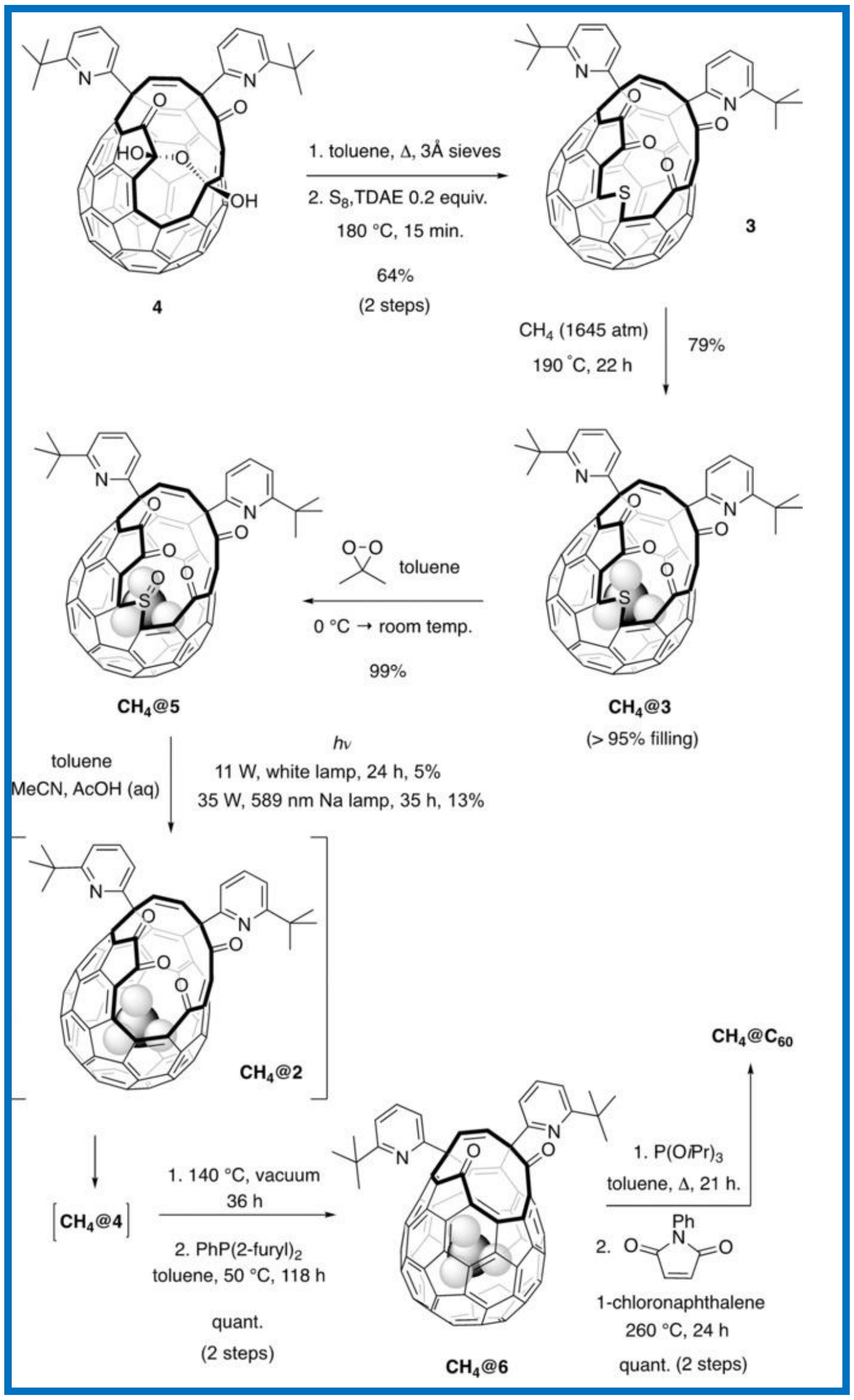

**Figure 19.** Synthesis of $CH_4@C_{60}$. Optimized $CH_4$ encapsulation by **3** successful closure sequences, including photochemical desulfinylation, applied to the first synthesis of $CH_4@C_{60}$. Reproduced from ref [174]. Copyright 2019, Wiley Online Library.

### 3.2 1D Carbon nanotube

The second class of carbon nano allotropes adsorbents, "Carbon Nanotubes" (CNTs) were discovered independently in 1993 by Iijima and Ichihashi[196]. CNTs generally are a wide range of carbon-made nanoscale diameter tubes. No matter the shape of the carbon tubular structure, CNTs are usually developed based on a hexagonal lattice of $sp^2$ carbon atoms, just like graphene sheets [169]. The edge of these sheets becomes a curve to form a final cylindrical tube with a high aspect ratio (i.e. length-to-diameter ratio). From the structural point of view, CTNs are compared to ‘‘rolled up’’ one-atom-thick sheets of graphene. Due to the difference in size and shape of CNTs and $C_{60}$, there is a significant difference between their inherent properties.

The most elementary form of CNT is a single graphenic wall with a closed cap at both end sides (Figure 20). The next type of CNTs member is single-walled carbon nanotubes (SWNTs), which are hollow tubes with about 0.4 to 2 nm diameter and several micrometers long. Furthermore, CNTs may be configured in double-walled (DWNTs) or multi-walled (MWNTs) carbon nanotubes based on the number of graphenic layers in the walls of the cylindrical structure [169]. Besides CNTs, other 1D carbon nanostructures such as carbon nanohorns [197, 198] and carbon nanoscrolls [199] have also been reported as potential adsorbents for gas storage (see Figure 20).

Although various CNTs hold the same basic structure as graphene, they are prepared entirely differently. CNTs are produced via several well-established methods, including a simple high-temperature (> 800 °C) synthesis using arc discharge or laser ablation and novel low-temperature (< 800 °C) techniques such as chemical vapour deposition (CVD) approach [200]. The CVD technique benefits from precisely controlling CNT's important parameters such as orientation, alignment, nanotube length, diameter, purity, and density.

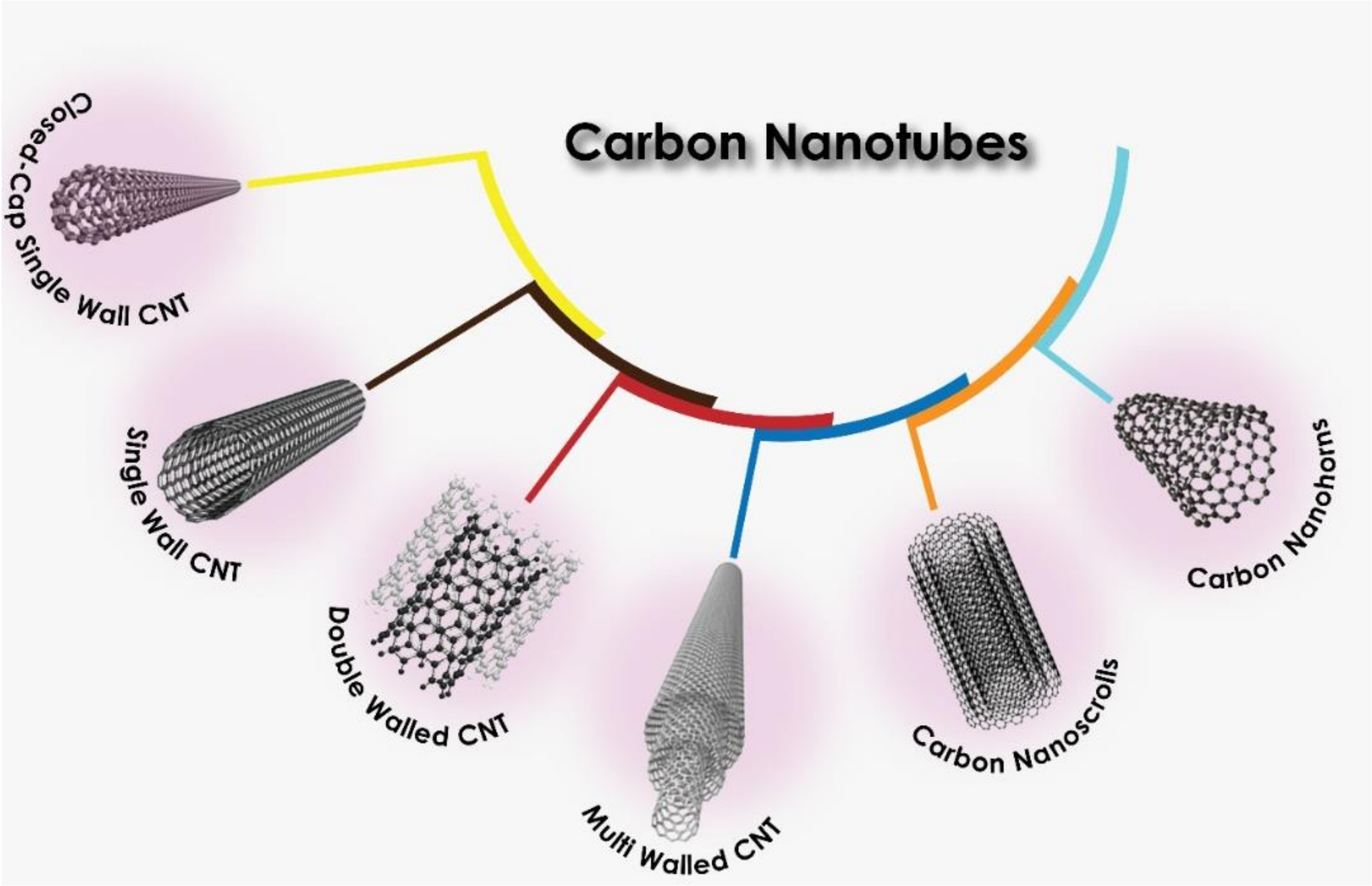

**Figure 20.** Different types of carbon nanotubes

Several attempts have been made to examine the impact of structural properties of curved graphenic surfaces for gas adsorption requests [201-203]. Bekyarova and colleagues fabricated open single-wall carbon nanohorns (SWNHs) as promising candidates for $CH_4$ storage through laser-ablating of graphite without a catalyst at room temperature [201]. The characterization of the prepared sample by TEM analysis and Raman spectroscopy showed a significant distortion in the structure of the compressed SWNHs. Such a disorderliness in the nanostructured carbons provided potential spots for storing $CH_4$ molecules. Thus, the obtained compound with high purity (~95%) and specific morphology illustrated a high ability to adsorb $CH_4$ (160 $cm^3.cm^{-3}$) at 35 bar and 30 °C.

Delavar et al. investigated the $CH_4$ adsorption capacity of two different types of MWCNTs adsorbents experimentally [202]. 'Type I' samples were prepared with the CVD method, with a purity of about 90%, the specific surface area of 126 $m^2.g^{-1}$, a mean pore diameter of 16 nm and a total pore volume of 0.5140 $cm^3.g^{-1}$. With 95% purity, the other prepared sample (Type II) was purchased from Alpha Nanotechnologies Company in China. The second sample possessed an outside diameter of 15-20 nm, inside diameter of ~4 nm, length of 30 mm, the specific surface area of 294 $m^2.g^{-1}$ and a total pore volume of 0.6231 $cm^3.g^{-1}$. The performance of MWCNT Type I with the developed mesoporous structure was not quite satisfactory for $CH_4$ adsorption (4.5 $mmol.g^{-1}$ @ 50 bar and 10 °C). In contrast, the MWCNT type II stored a considerably significant amount of $CH_4$ (33 $mmol.g^{-1}$) at the same condition. Such an exceptional $CH_4$ storage capacity can be justified by the unique surface properties of Type II MWCTs such as their highly uniform pore size and high pore volume.

Numerous studies have suggested that defects in CNTs structures increase the possibility of their gas storage capacity [201, 204, 205]. Defects or disorder cavities on the CNT framework may develop the surface area and pore volume of CNTs, act as entry points for gases, and shorten the diffusion pathway. Orimo and coworkers proposed the technique of “mechanical milling” to promote the hydrogen uptake of a nanostructure materials [206]. In a similar study, Liu and co-workers observed that the hydrogen loading was enhanced effectively as the milling time increased [207]. Moreover, some authors proposed the “post-activation” treatment for improving the porosity of CNTs and enhancing the defection and irregularity inside the nanotube structure [97, 208]. Adeniran and Mokaya made a series of carbon nanotubes at low temperature (180 °C) using carbon tetrachloride as precursor and ferrocene/Ni as substrate/catalyst [97]. The obtained CNTs showed a predominantly amorphous framework with large tube diameters of 180–300 nm, a wall thickness of *ca*. 25 nm, a surface area of 470 $m^2.g^{-1}$ and pore volume of 0.39 $cm^3.g^{-1}$. After KOH-activation, the surface area and micropore volume were increased to 1479–3802 $m^2.g^{-1}$ and 1.65-2.98 $cm^3.g^{-1}$, respectively. They reported that the activated-CTN with KOH/carbon ratio of 4, contained large micropores and a significant proportion of mesopores and exhibited hydrogen storage of up to 7.5 wt%, 9.7 wt.% and 14.9 wt.% at -196 °C and 20, 40 and 150 bar, respectively. Due to practical difficulties associated with experiments in controlling/adjusting the geometrical properties of CNTs, the reported results in the literature are often controversial. One best way to precisely clarify the impact of the design parameters on the gas storage capacity of carbon nanotubes is using theoretical molecular studies. This method makes it possible to describe the nature of the interactions between gas and host nanomaterials, which helps to better understand the basic mechanisms of gas storage. As an example, the physisorption of pure $CH_4$ on bundles of single-wall carbon nanotubes (SWCNT) was studied by computational methods to investigate the influence of the material's porosity on the storage capacity and the adsorbate distribution [209]. In this research, the concept of CNT porosity was considered in two ways based on different (i) diameters of the nanotube inner cavity and (ii) the volume of the interstitial channels. The maximum load capacity was observed for the selected nanotubes with a smaller diameter. The small inner cavities efficiently trap $CH_4$ molecules and present a larger van der Waals (vdW) stabilization energy. The simulation results also proved that there is a significant chance of improving SWCNT-bundles' load capacity if the gap between the tubes can be controlled accurately.

In another theoretical study, Peng et al. performed a molecular simulation on $CH_4$ and $CO_2$ storage/separation on carbon nanoscrolls [199]. They analysed the effects of temperature and pressure, interlayer spacing, vdW gap and innermost radius systematically on the gas storage of these curve-shaped nanostructured materials. It was found that the adsorption of gases on pristine nanoscrolls is relatively low. However, both $CH_4$ and $CO_2$ adsorption capacities experience a significant improvement after expanding the interlayer spacing (see Figure 21). In similar research, Mpourmpakis and co-workers performed a multiscale theoretical approach to investigate the hydrogen storage capacity of recently synthesized carbon nanoscrolls [210]. They found that pure carbon nanoscrolls cannot properly accumulate hydrogen because of their small interlayer distance. However, opening the spiral structure to approximately 0.7 nm followed by alkali doping can make them auspicious materials for hydrogen storage application, reaching 3 wt.% at ambient temperature and pressure (Figure 22).

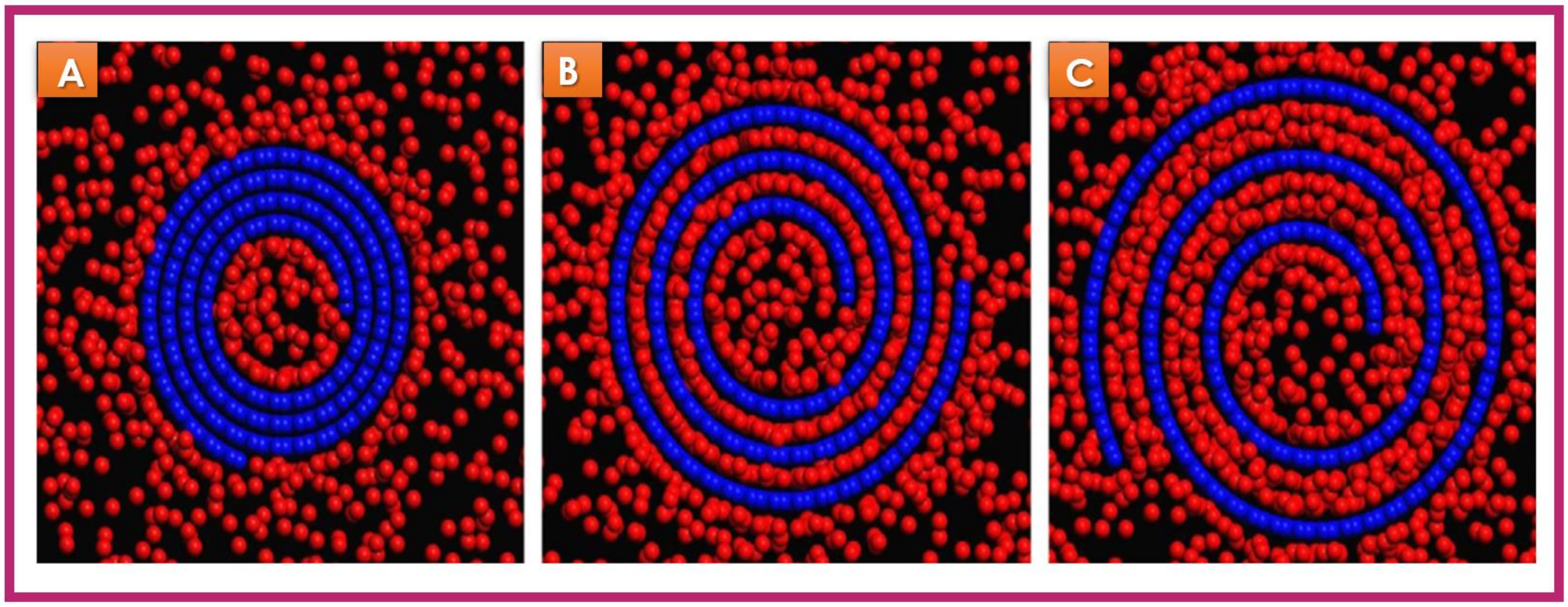

**Figure 21**. Snapshots of $CH_4$ adsorbed on isolated carbon nanoscrolls at 25 °C and 90 bar: (A) D = 0.34 nm; (B) D = 0.7 nm; (C) D = 1.1 nm. The blue and red spheres denote carbon atoms on nanoscrolls and $CH_4$ molecules, respectively. Reproduced from ref [199]. Copyright 2010, Elsevier Ltd.

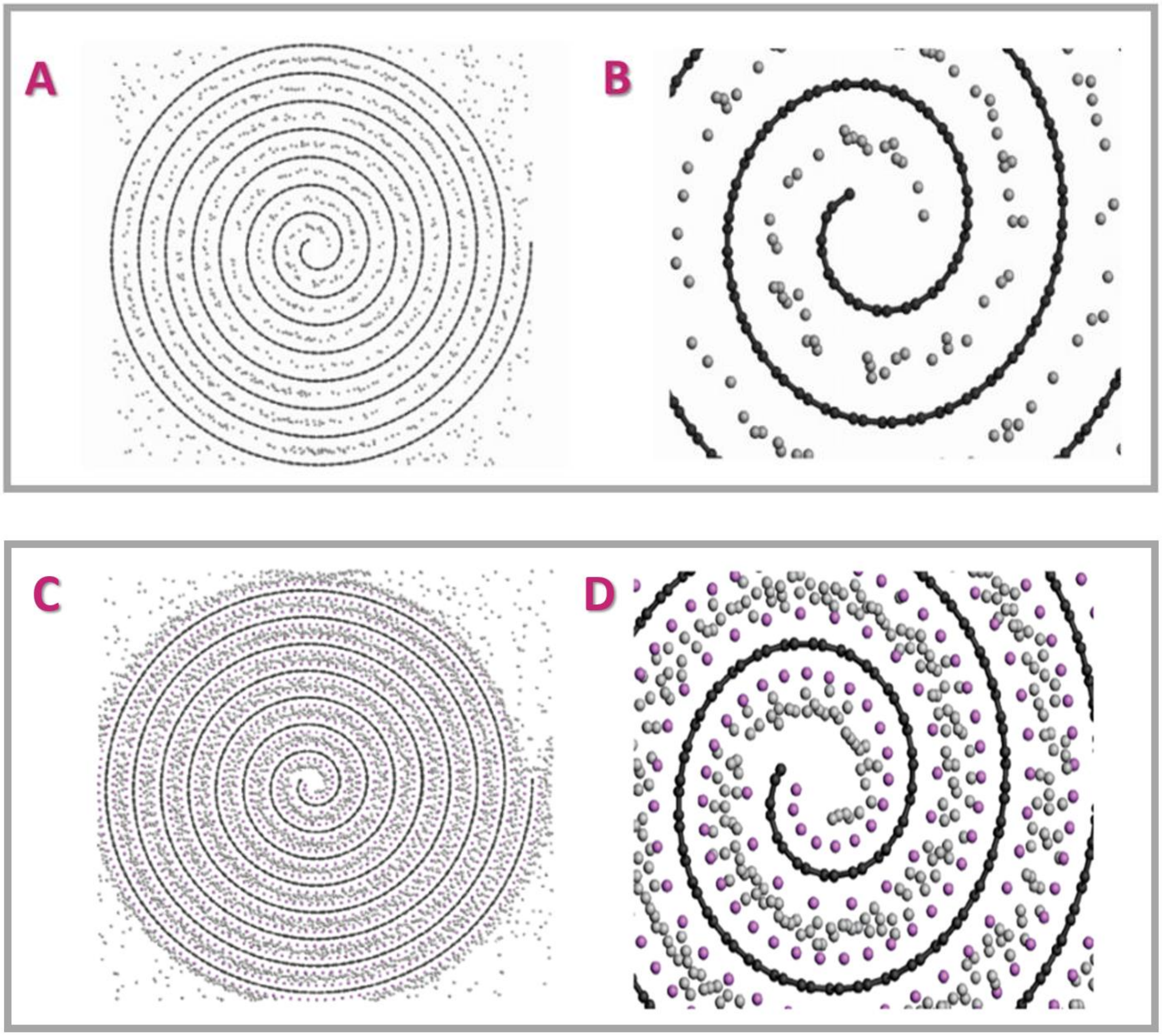

**Figure 22.** Snapshots from GCMC simulations of hydrogen adsorption on undoped CNS structures of (A) 0.7 nm interlayer distance and (B) zooming of the central part at room temperature and 100 bar pressure. The GCMC simulation of hydrogen adsorption on (C) alkali doping of CNS structures with 0.7 nm interlayer distance and (D) zooming of the central part at room temperature and 100 bar pressure. Reproduced from ref [210]. Copyright 2007, American Chemical Society.

The configuration arrangement of CTNs affects the gas uptake characteristics of these nanomaterials. In other words, the gas adsorption in interstices of nanotube arrays plays a significant role in their total volumetric and gravimetric gas capacity. A grand canonical Monte Carlo (GCMC) simulation was performed to optimize an array of armchair SWCNTs arrangments in a triangular shape for gas storage [211]. The results further confirmed that the interstitial adsorption highly depends on the van der Waals gap between the tubular arrays. From a theoretical point of view, the interstitial adsorption process is tricky and varies with the size of tubes in addition to the distance between the tubes in a triangular array [212]. It has been demonstrated in Figure 23 that a triangular array of (14,14) nanotubes with a diameter of 1.9 nm separated by a vdW distance of 0.34 nm, possesses the volumetric storage capacity of up to 173 $V.V^{-1}$ at 35 bar and 298 K, which is 96% of the landmark value 180 $V.V^{-1}$ introduced as DOE target in 2015. Zhang and Wang used GCMC simulations and DFT calculations to explore the adsorption of $CH_4$ in a square lattice of nanotubes array with different diameters of 2.04-4.077 nm, wherein the tubes were separated by a van der Waals distance of 0.334 nm [213]. They found that an array of nanotubes with diameters of 4.077 nm can store about 22 $mmol.g^{-1}$ (35 wt %) at 300 K and 60 bar. Most of the reported values from theoretical studies seem promising for gas storage in CNTs (Table 7). Still, experimental attempts to reach or even approach such values in this class of nanomaterials have been unsuccessful. Under actual conditions, none of the carbon nanotubes synthesized have exhibited such ideal storage values. The conflicts between theory and experiments are most probably due to the fact that the nanotubes used in experiments are far from ideal theoretical structures. The synthesized CNTs are often distorted, contain mixtures of opened and unopened and single-walled and multiwalled nanotubes of various diameters and helicities [214].

**Table 7**. Gas storage properties of CNTs designed by theoretical and experimental study.

| Adsorbent | Study Condition | $S_{BET}$ ($m^2.g^{-1}$) | $H_2$ Uptake (wt.%) | $CH_4$ Storage Capacity ($V.V^{-1}$) | Selective Separation Capacity (SSC) | | Measurement Condition | Reference |
|---|---|---|---|---|---|---|---|---|
| | | | | | $CO_2/CH_4$ | $CH_4/N_2$ | | |
| Commercial MWCNT | Experiment | > 500 | - | 0.98 $mmol.g^{-1}$ | 2.39 | - | $CH_4$ @ 30 °C, 3.5 bar<br>SSC@ 30 °C , 1 bar | Molyanyan et al.[203] (2016) |
| SWCNT with chirality (10,0) | Dynamics simulations with DLPOLY method | - | - | 3.9 | - | - | $CH_4$ @ 0 °C , 35 bar | Vela et al.[209] (2011) |
| Isolated carbon nanoscrolls | Molecular GCMC & GEMC simulations | - | - | 262.70 | 1.57 | - | $CH_4$ @ 25 °C , 65 bar<br>SSC@ 25 °C, 60 bar | Peng et al. [199] (2010) |
| Commercial MWCNT | Experiment | 126 | - | 4.02 $mmol.g^{-1}$ | - | - | $CH_4$ @ 25 °C, 50 bar | Delavar et al.[202] (2014) |
| MWCNT | Prepared by CVD method | 294 | - | 28.44 $mmol.g^{-1}$ | - | - | $CH_4$ @ 25 °C, 50 bar | Delavar et al.[152] (2014) |
| SWNHs | Prepared by laser-ablating graphite | 1030 | - | 96 | - | - | $CH_4$ @ 30 °C, 35 bar | Bekyarova et al.[201] (2003) |
| CNT | Multiscale theoretical approach based on DFT calculation | - | 5.24 | - | - | - | $H_2$ @ -196 °C, 10 bar | Mpourmpakis et al.[210] (2007) |
| Scroll | Multiscale theoretical approach based on DFT calculation | - | 3.08 | - | - | - | $H_2$ @ -196 °C, 10 bar | Mpourmpakis et al. [160] (2007) |
| CNT | GCMC simulations | 759 | 0.44 $mmol.g^{-1}$ | 2 $mmol.g^{-1}$ | - | - | $H_2$ @ 25 °C, 25 bar<br>$CH_4$ @ 25 °C, 25 bar | Kumar et al.[215] (2012) |
| CNT4700 | Prepared by hydrothermal method | 3202 | 6.7 | - | - | - | $H_2$ @ -196 °C, 20 bar | Adeniran & Mokaya[112] (2015) |

### 3.3 2D Graphene

Another important group of graphitic nanostructures is the building block of graphite known as "Graphene". Graphene is an ideal one-atom-thick two-dimensional (2D) $sp^2$ bonded carbon sheet arranged in a hexagonal honeycomb lattice [169]. Although the structure of graphene had been experimentally identified in 1962 by Boehm et al [216] and predicted decades ago [217], graphene was isolated and characterized as a 2D carbon nano-allotrope for the first time by the efforts of Novoselov and Geim in 2004 [218]. About six years later, in 2010, Novoselov and Geim received the Nobel prize in physics for detecting such a fantastic carbon-based material [169].

These particular 2D carbon nano-allotropes have attracted interest in various research areas such as medicine, aviation, as well as automotive and aerospace industry, and military applications because of their unique structure and exceptional characteristics. Recently, a significant number of investigations have been focused on applying graphene and its oxygen-containing derivatives, so-called graphene oxide (GO), for gas storage/separation requests. These carbonaceous structures usually possess a large surface area with adjustable microstructures, excellent electronic conductivity, and remarkable mechanical, chemical, thermal, and electrochemical stability.

The graphene's quality is affected strongly by its production method. Briefly, the fabrication technics can be classified as chemical vapour deposition (CVD), plasma-enhanced CVD, graphitization, solvothermal, thermal and chemical exfoliation [219]. Generally, high-quality graphene is made of more expensive procedures (i.g. graphitization and exfoliation), while affordable synthesis technics (e.g. chemical vapour deposition and solvothermal) result in materials with poor electrical conductivity or mechanical strength. A few recently published review articles reported more details on graphene-based adsorbent synthesis procedures [219-221].

In relevant research, the hydrogen storage capacity of the so-called “pristine single-layer graphene sheet” with a BET specific surface area of 156 $m^2.g^{-1}$, was evaluated [222]. The results showed that the $H_2$ adsorption is about 0.4 wt.% (at -196 °C and 1 bar) and 0.2 wt.% (at 25 °C and 60 bar), respectively. As it can be seen, the hydrogen adsorption capacities are far below the DOE target expectation of 5.5 wt.%. Therefore, attempts have been made to improve the graphene nanosheet's gas adsorption capacity via different modification approaches [223-225]. Post chemical/physical activation of graphene sheets seems to be a successful approach to improving gas capacity. Using a facile and scalable synthetic approach, Huang et al. fabricated graphene nanosheets with hierarchically porous structures [226]. The produced steam-activated graphene sheets showed appropriate interconnected micro-meso-macroporous structures with high specific surface areas up to 1238 $m^2.g^{-1,}$ and ultra-large pore volumes up to 4.28 $cm^3.g^{-1}$. These materials showed a high hydrogen storage capacity of 1.47 wt.% at -196 °C and 1 bar. Klechikov et al. optimized a post-KOH-activation procedure to enhance the graphene scaffold's porosity and gas adsorption properties (Figure 24) [227]. As-prepared adsorbent showed high BET surface area (values up to 3400 $m^2.g^{-1}$) and pore volume up to 2.2 $cm^3.g^{-1}$. The maximum hydrogen uptake of ~7 wt.% (at -196 °C and 129 bar) was observed for these activated graphenes.

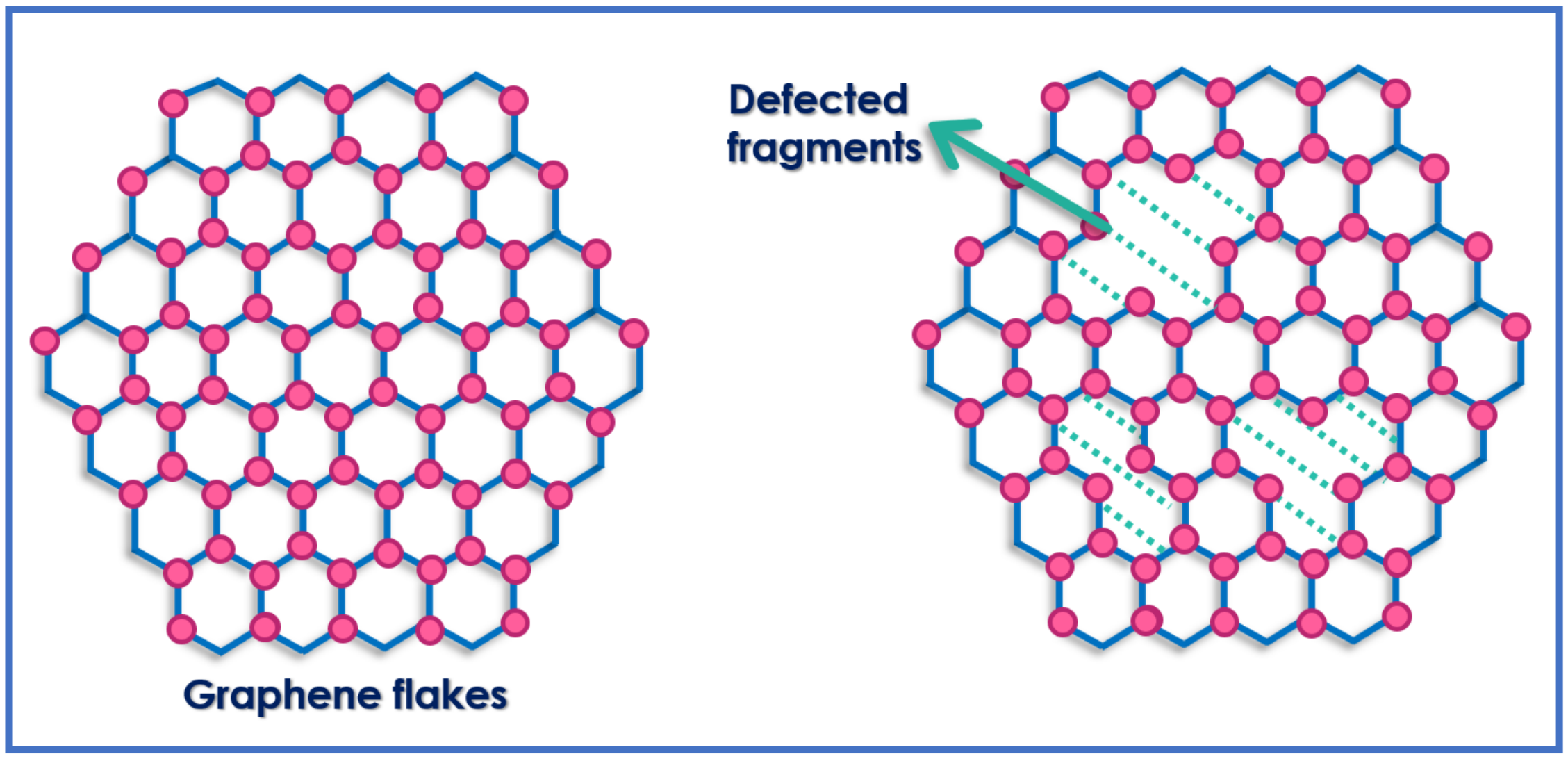

**Figure 24.** Schematic representation of possible defected fragments of graphene flakes.

A series of porous graphene-based adsorbents were fabricated by Srinivas et al. using KOH-activation of thermally exfoliated GO (exf-GO) and solvothermal reduced GO (rGO) precursors [228]. They reported an efficient approach to significantly increasing the graphene oxide surface area from 10 $m^2.g^{-1}$ up to about 1900 $m^2.g^{-1}$. Besides, a linear trend between the BET surface area of graphene-based adsorbents and their $CH_4$ adsorption capacity was observed. The optimum activation condition of 800°C and GO(1): KOH(9) ratio was reported to obtain a sample with hierarchical pore structure (1.65 $cm^3.g^{-1}$) and high $CH_4$ storage capacity (0.175 $mg.g^{-1}$). The feasibility of layer spacing and adding dopants on multilayer graphene sheets to improve their $CH_4$ adsorption was theoretically investigated in another study [229]. As depicted in Figure 25, simulation results indicated that the number of adsorbed $CH_4$ molecules increases with the expansion of the graphene layer gap. However, the volumetric storage capacity might be reduced due to the increase in the volume space.

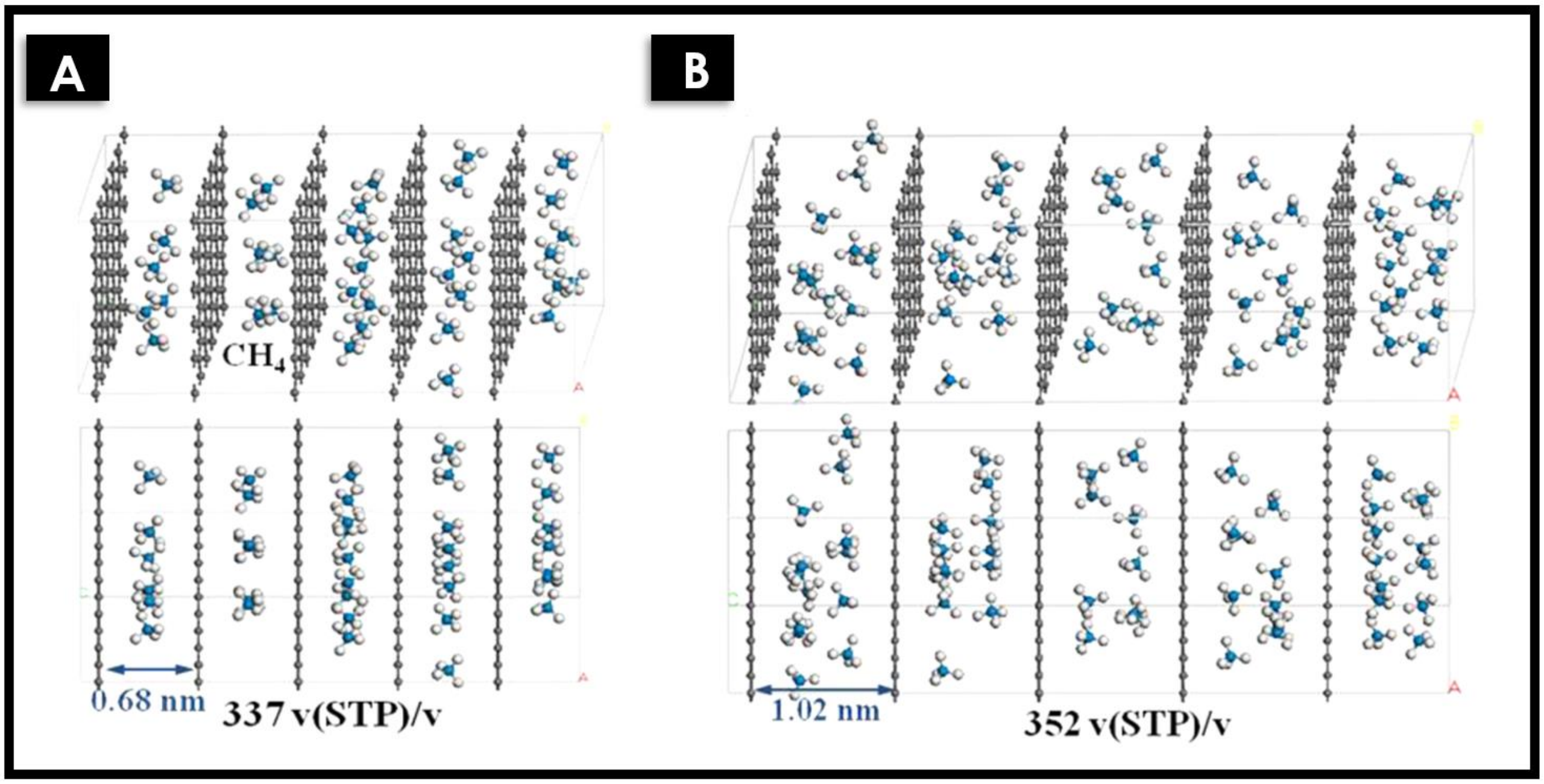

**Figure 25.** Snapshots and adsorption capacities of $CH_4$ in MGN with a dGL of (A) 0.68 nm and (B) 1.02 nm at 25 °C and 100 bar, including the three-dimensional view and front view. Reproduced from ref [229]. Copyright 2012, American Chemical Society.

The reported $CH_4$ and $H_2$ capacity of graphene sheets is well below the recent DOE targets in either of the theoretical and experimental studies. As mentioned, the linear relationship between the porous textural properties and gas adsorption capacity of the graphene sheet was reported in the literature. Theoretical studies showed that ideally, one particular graphene sheet can have a BET surface area as high as 2630 $m^2.g^{-1}$, which means about 1315 $m^2.g^{-1}$ for each side of the sheet [230]. One best solution to fully exploiting the gas storage properties of 2D graphene sheets is to generate three-dimensional structures by introducing “spacers” among the graphene layers and producing interconnected defected graphene sheets [221, 231, 232]. The following section presents progress on the fabrication of 3D graphene-based structures for gas adsorption purposes.

## 3.4 3D Graphene

When it comes to practical applications, 2D graphene sheets might lose their inherent properties and consequently show poor performances far below the theoretical targets. One effective way to promote graphene sheets' processability and practical application in energy storage is to convert them into well-organized and interconnected three-dimensional (3D) structures made up of more than ten stacked layers of graphene. With such a solution, the exceptional properties of 2D graphene materials will be preserved, and the obtained 3D samples can be easily used without the concern of restacking.

Accordingly, the last but not the least member of the large family of carbon nano-allotropes is defined as a graphitic framework, where graphene layers weakly interact through van der Waals (vdW) forces. Different classifications of 3D graphene architecture have been presented since 2009, including pillared graphene frameworks (PGFs) [233], graphene oxide frameworks (GOFs) [234], multilayer graphene frameworks (MGFs) [221], graphene foams [235], and graphene sponges [236]. Different preparation methods of hydrothermal reduction [231, 237], chemical reduction [238, 239], laser casting [240, 241], 3D printing [242, 243], chemical vapour deposition (CVD) [221, 244] and on-site polymerization processes[245, 246] have been reported to produce these carbonaceous adsorbents.

Different types of 3D nanoporous graphene structures have been screened computationally and experimentally to find adsorbents with capacity close to the DOE target for $CH_4$ and $H_2$ storage [234, 247-249]. Hassani et al. employed hybrid molecular dynamics simulations to investigate the $CH_4$ storage capacity of nanostructure adsorbents of pillared graphene frameworks (PGFs) [247]. These frameworks are composed of parallel multi-layers of graphene, connected by vertical organic linkers (so-called holders). They found the $CH_4$ uptake of the PGFs is about 22% higher than the pristine graphene sheets. In another study, the impact of length of linker inter-space on the $CH_4$ adsorption capacity of hypothetical graphene oxide framework was assessed by simulation [234]. Results showed that neither large linker interspace of 1.4 nm nor small linker interspace of 0.8 nm, but an optimum linker interspace distance (1.1 nm) results in the highest $CH_4$ adsorption capacity (19.70 $mmol.g^{-1}$ at 25 °C and 60 bar).

The highest volumetric $CH_4$ uptake value of 317 $V.V^{-1}$ was reported in a simulation study of general multilayer frameworks (GMFs), multilayer graphene frameworks (MGFs) and pillared graphene frameworks [248]. The stored gas amount reported for these 3D graphene-based structures remarkably can meet the latest value of the DOE target. In an experimental study, a novel, cost-effective and single-step technique was reported for mass production of 3D porous graphenes, using bagasse as a source of carbon [249]. In this research, bagasse was chemically activated using KOH at 850 °C with a ratio of 1(bagasse):6(KOH). The final microporous adsorbent showed large micropore volume of 1.45 $cm^3.g^{-1}$, an unprecedented surface area of ~3000 $m^2.g^{-1}$ and a promising methane storage capacity of 193 $V.V^{-1}$ at 25 °C and 35 bar.

Recent advances in hydrogen storage on 3D graphene materials confirm their possible high capacity and acceptable stability for practical usage. In a simulation study, using idealised models, Garberoglio et al. assessed the amount of $H_2$ uptake of hypothetical organic pillared reduced-graphene-oxide sheets [169]. It was found that two important factors of (i) the density of the pillars and also (ii) the corrugation of graphene sheets have a profound impact on the gas adsorption properties of these materials. Another approach was adopted by Kumar and colleagues, where the $H_2$ storage of two different porous graphene frameworks (PGFs) was investigated via theoretical and experimental assessment [250]. Both of the Three-dimensional frameworks exhibited high surface area (825 $m^2.g^{-1}$), pore volumes (0.74 $cm^3.g^{-1}$) and $H_2$ uptake (1.2 wt.% at -196 °C and 1 bar) due to the effect of pillaring.

Macroscopic 3D graphene sponges and foams are also attractive materials for gas storage/separation. Lyth and coworkers reported a low-cost and scalable technique for fabricating 3D graphene foams through the combustion of sodium ethoxide [251]. As-prepared graphene foams showed a surface area of 1296 $m^2.g^{-1}$ and $H_2$ storage capacity of 2.1 wt.% at -196 °C and 10 bar pressure. The prepared 3D graphene foams also showed a higher $H_2$ uptake value than the commercially obtained graphene sample (1.2 wt.% at -196 °C and 10 bar), which might be attributed to the higher surface area of the graphene foam. In another study, nanoporous spongy graphene oxides were synthesized using combined wet chemical reduction and freeze-drying techniques [252]. Potential application of these spongy graphenes for the gas storage/separation uses was analyzed by $H_2$, $CO_2$ and $CH_4$ sorption measurements at different temperatures. The measured gas storage capacity of the samples at 1 bar was about 0.5 wt.% for hydrogen (-196 °C), 3.7 wt.% for carbon dioxide (0 °C) and 0.4 wt.% for $CH_4$ (0 °C). Moreover, the 3D sponges showed high $CO_2$ over $CH_4$ selectivity of 95:1 at 0 °C and 0.7 bar pressure.

## 4. Summary and outlook

Considering the energy crisis and environmental pollution as the current global threats, developing practical, clean, affordable, and available energy storage systems becomes vital. High-pressure storage of energy carrier gases such as $CH_4$ and $H_2$, as adsorbed phase on porous materials, is a promising safe alternative for current expensive and energy-consuming techniques. The porous carbons family, particularly activated carbons, has a long usage history in energy storage systems. These tunable structures, with unique inherent properties, are experiencing continual development to satisfy the requirements of emergent applications. This paper reviewed various recent design strategies to synthesise carbonaceous adsorbents for $CH_4$ and $H_2$ storage.

Carbon-based adsorbents in different forms of granule, powder, fibres, foam, pellet and monoliths, with a high surface area and good packing density, can play a meaningful role in gas storage/separation applications. The key factor for gravimetric gas uptake of these adsorbents is their porosity; however, for the volumetric gas storage capacity, another critical parameter of packing density should be counted simultaneously. Between these two different storage capacity scales, the volumetric uptake of energy carriers is more important and applicable for onboard vehicular applications. In the case of methane storage, the highest volumetric uptake can reach up to 200 $cm^3.cm^{-3}$ at 25 °C and 65 bar, getting closer to or meeting the new DOE targets. For hydrogen storage, the adsorbent with a well-developed microporosity exhibits high storage capacity of 4 wt.% at -196 °C and 1 bar condition.

Traditional activation of carbon materials enlarges the pores to prepare adsorbents with high surface area. However, to better control and tune the porosity, new techniques such as HTC, templating, and CDC methods have been developed to fabricate carbonaceous materials. These methods showed repeatable results, and products with relatively well-controlled structural characteristics (i.e. surface area, pore-volume, pore size, particle size, morphology, etc.), and, as a result, higher gas uptake. For example, some zeolite-templated carbons showed to have a high pore volume, fined-tuned PSD and exceptionally high methane storage of up to 210 $cm^3.cm^{-3}$ at 25 °C and 65 bar.

Activated carbons are recognized as well-studied adsorbents in many application fields. Still, limitations such as their relatively low gas storage efficiency, disorder pore geomorphology, and even etching contamination during the activation procedure are associated with these adsorbents.

It is essential to explore new configurations of carbons with fewer inherent problems to overcome the limitation of these valuable materials. Over the past decade, carbon nano-allotrope adsorbents with highly ordered organized structures were introduced to compete with conventional amorphous disorder activated carbons. These novel adsorbents with different classes of zero (0D, carbon fullerenes), one (1D, carbon nanotubes), two (2D, graphenes) and three (3D, graphitic carbons) dimensions show great potential to serve as a suitable platform for loading gases such as methane and hydrogen. There is a strong possibility of significantly improving gas adsorption on carbon nano-allotropes structures due to defects inside their unique carbon cell. Recent studies on carbon nano-allotropes have revealed that the gas storage capacity of these materials can be enhanced by treating them with acids/basic media under different thermal conditions. However, most of the promising reported gas storage on this class of nanomaterials is based on theoretical studies. It seems experimental attempts to reach or even approach such values have been unsuccessful and need further investigation.

In the end, it should be acknowledged that despite the latest progress on developing novel materials, activated carbons remain a primary option for the fabrication of gas molecular adsorbent due to their accessibility, low cost, and ability to mass production and facile preparation methods. However, as motioned earlier, these materials also associated with some drawbacks and further technological and scientific research is required to improve carbon-based adsorbents:

For instance, new-faced materials like hybrid nanocomposite structures can be used as an energy-efficient platform for advancing porous carbon sorbents. One advantage of using such novel structures' is to benefit the unique properties of its two (or even more) pristine components at the same time, however, more precise research and study should be done to economize and facile the fabrication methods of these adsorbents. Furthermore, the exact mechanism of gas adsorption on, inside or outside, curved-carbon structured or pillared graphene adsorbents is still controversial. Thus, further investigation should be performed to find fundamental molecular gas adsorption mechanisms inside the skeleton or on the surface of these carbon structures as next-generation gas adsorbers. Finally, we hope the topic of carbon-based materials continuously develops in other emerging regimes and attracts more scientists to join.